# DISTANCES AND ABSOLUTE MAGNITUDES OF DWARF NOVAE:

# MURMURS OF PERIOD BOUNCE

by

Joseph Patterson[1]
Department of Astronomy
Columbia University
13 October 2010

## ABSTRACT

We redetermine the relationship between absolute magnitude and orbital period for dwarf novae, based on 46 stars with good distance estimates. This improves upon Warner's previous relation, building upon today's improved estimates of distance and binary inclination, and greater wavelength coverage. Together with other distance and dynamical constraints, this calibration is then applied to a set of ~300 known or likely dwarf novae of short orbital period, to study the dependence of quiescent $M_v$, time-averaged $M_v$, mass ratio $q$, and white-dwarf temperature $T_{WD}$, on $P_{orb}$. These distributions show that stars become much fainter as they approach minimum $P_{orb}$, and appear to show **evolutionary tracks** as the secondary is whittled down by mass loss. Stars on the lower branch have the expected properties of "period bouncers" – with a feeble secondary, faint accretion light, cool white dwarf, and long recurrence time between eruptions. Period bounce seems to occur at a mass of 0.058±0.008 M$_o$. Stars on the lower branch may also have higher velocities and heights above the Galactic plane, consistent with a greater age. Some are very nearby, despite strong selection effects discriminating against the discovery of these faint binaries accreting at very low rates. Period bouncers appear to be very common, and probably would dominate a complete census of cataclysmic variables.

> When you got nothing, you got nothing to lose
> You're invisible now...
>                                           – Dylan (1965)

---

1  Department of Astronomy, Columbia University, 550 West 120th Street, New York, NY 10027; jop@astro.columbia.edu

## 1. INTRODUCTION

Distance is the *sine qua non* of astrophysics. A distance estimate is required to convert from flux to luminosity, and stellar physics is about luminosity, not flux. Unfortunately, distances to cataclysmic variables (CVs) are particularly difficult to estimate, because the dominant light source is not a star but an accretion disk – preventing straightforward application of physical methods developed for single stars. But clues from geometry (parallax and motions) are still available, and Kraft & Luyten (1965) used these clues on a few of the brightest stars to make rough estimates for dwarf novae, the most common type of CV: $M_v$ = +4.5 in eruption, and $M_v$ = +7.5 in quiescence. Warner (1987) considered more stars (25) and substantially improved the estimate for erupting dwarf novae at maximum light: $M_v$ = +5.64 – 0.26 $P_{orb}$, where $P_{orb}$ is the binary's orbital period in hours.

Theoretical studies (Smak 1989, 2000; Cannizzo 1998; Osaki 1996) have provided a simple and compelling interpretation of this result. The eruption tends to be a standard candle because every accretion disk "blows" when its surface density and resultant temperature rise to ~8000 K, where hydrogen opacity rapidly increases and a transition to a "hot disk" state occurs. To a first approximation, we would then expect all freshly erupted disks to be equally bright – basically radiating as optically thick surfaces with $T$ = 20000 K, where the heating wave is quenched because the whole disk is in the hot state and radiating. But binaries of longer $P_{orb}$ have larger disks, and therefore should be brighter.

Over the years, this relation has done yeoman service as a rough distance estimator for all erupting dwarf novae. Harrison et al. (2004, hereafter H04) tried to improve upon it, with high-quality parallaxes and a careful analysis of empirical data on the eruptions. But their relation is based on only 6 stars, with a scatter among them of ~0.6 mag, exceeding the difference between their relation and Warner's.

In 2010, many more ingredients are present for distance estimation: more parallaxes, proper motions, and much better parsing of a CV's light into its components: disk, white dwarf, and secondary star. In this paper we use these clues for an improved $M_v(P_{orb})$ relation for dwarf novae at maximum light, based on 46 stars with acceptable constraints on distance and binary inclination. This relation is very similar to that of Warner, underlining the basic truth that *dwarf nova eruptions are a pretty good standard candle*. We then use this calibration to explore the dependence of quiescent, and time-averaged, $M_v$ on $P_{orb}$ for ~300 short-period CVs which are confirmed and likely dwarf novae. We also study the dependence of mass ratio and white-dwarf temperature on $P_{orb}$, and explore the issues of recurrence time, scale height, and space density. The results appear to reveal a large population of "period bouncers". Most of these stars probably still elude discovery; but their high representation in the solar neighborhood suggests that they may well be the majority species of CVs.

## 2. PRIMARY DISTANCE MEASURES

The distances of nearby CVs are roughly 100 pc, which is at the edge of what is feasible with trigonometric parallax measures from photographic plates. As discussed by Thorstensen (2003, hereafter T03), the early photographic parallaxes have mostly proved incorrect. With digital measures, accuracy has improved by a factor 3–5, or even greater for stars with the tiny image profiles available from space (HST, Hipparcos). So we now have good parallax constraints for a few dozen CVs (T03; H04; Thorstensen, Lepine, & Shara 2008, hereafter T08); these are certainly the best distance measurements available.

Also useful are techniques which provide basically a *photometric parallax* of the secondary. At quiescence, K or M secondaries are often prominent in the spectrum. Roche geometry specifies the secondary's emitting area, and the spectral type specifies the surface brightness; so the observed flux can be compared, yielding a distance. This is sometimes called the "Bailey method", especially in the simple form where the *K*-band brightness alone is used (Bailey 1981). This was the primary basis of distance estimation in Warner's 1987 study. We now have much better measures of the secondary's $T_{eff}$, spectral type, and *K* magnitude, partly due to improved results from red/infrared spectroscopy (e.g. Friend et al. 1990; Dhillon et al. 2000; Ishioka, Sekiguchi, & Maehara 2007), and partly due to the 2MASS *JHK* survey (Skrutskie et al. 2006). Beuermann (2006) improved the Bailey relation by adopting a better calibration between spectral type and surface brightness. His figures vividly illustrate the importance of this correction, and the relation can be further improved by using a more accurate mass-radius relation, empirically derived from CVs themselves, rather than model stars (Patterson et al. 2005a, hereafter P05; Knigge 2006, hereafter K06).

A variation on the Bailey method is the "period-luminosity-color" (P-L-C) relation, which uses the infrared color to characterize a star's surface brightness. Ak et al. (2007, 2008) uses this method, after calibrating with stars of known trigonometric parallax. In principle, this can be a fine technique. But it has several serious drawbacks when applied to short-period dwarf novae:
(1) It gets confounded when the flux distributions are composite (white dwarf plus disk), which is usually true of a quiescent dwarf nova (DN).
(2) Its applicability to an optically thin disk (likely the main light source in a quiescent DN) is unknown, and not trustworthy since there is no real "surface".
(3) The main data source was the 2MASS survey, with 8-second exposures giving only very coarse measures of infrared color, or none at all, for the faint stars we study.
(4) Only 7 short-period stars are in the trig parallax calibration, and even that small supply probably represents three *different* types of light source dominating infrared emission (secondary, white dwarf, and disk).
These problems are presently too serious to use this technique on short-period dwarf novae.

A second simple photometric parallax uses the observed brightness of the white dwarf (WD). By 1985, the signature of 5–10 DA WDs had been recognized in the spectra or light curves of CVs (Smak 1984, Sion 1985, Patterson & Raymond 1985). From WD model atmospheres, broadband colors and line profiles yield (after removal of accretion light) estimates for $T_{eff}$, which were in the range 10000–18000 K. With a plausible estimate for the WD's radius, this yields $M_v$ and therefore a distance estimate. More such stars were identified as a result of the surveys of the 1990s, especially the SDSS, and more and better estimates for $T_{eff}$ were obtained by fitting Lyman-α profiles accessible with HST (see e.g. Gansicke et al. 2005, Szkody et al. 2007).[2]

White-dwarf estimates are no panacea, however. "Recognized in the spectra or light curves" is a fairly restrictive condition. Clear recognition usually requires either a sharp eclipse, or the WD's certifiable domination of the binary's blue (4000–5000 A) light. Both apply only to the most intrinsically faint CVs, where the disk does not interfere too much. Perhaps the most accurate of these constraints come from high-speed multicolor light curves of sharp eclipsers (e.g. Littlefair et al. 2008, hereafter L08). In particular, the *ugriz* or *UBVRI* fluxes of the WD can be compared to the theoretical fluxes of WDs; flux ratios yield the temperature, and Kepler's Laws plus the WD mass-radius relation yield accurate values of mass, radius, and binary inclination. When *V* light curves only are available, the WD's *V* flux can be compared to an interpolation in DA models (Bergeron, Wesemael, & Beauchamp 1995):

$$M_v = 11.24 - 0.12[T_{WD}/(1000 \text{ K}) - 15] + 2.5[(M_{WD}/M_\odot) - 0.6], \tag{1}$$

where $T_{WD}$ and $M_{WD}$ are the WD's temperature and mass. This approximation works well for WDs within hailing distance of typical values in short-period dwarf novae (15000 K, 0.8 $M_O$). It does, however, depend on the assumption that the "white dwarf" component is actually a fully visible WD photosphere – not obscured by structures in the disk, and not greatly affected by a thin and hot equatorial boundary layer.[3]

These three techniques are our primary methods of estimating distance. The accuracy of both methods of photometric parallax is enhanced by data over a broad wavelength range (about a decade), to permit an accurate parsing into the relevant components (secondary, WD, disk). The great telescopes/surveys of the 1990s (HST, 2MASS, SDSS) have given access to that wavelength range.

## 3. ORBITAL PERIOD, APPARENT MAGNITUDE, EXTINCTION, AND INCLINATION

Aside from the new data furnished by all these fancy telescopes, we need plenty of help from infantry: the orbital period, apparent visual magnitude at maximum light, visual extinction, and binary inclination. The quality and quantity of this data has greatly improved since Warner's 1987 study; details are discussed below, and refer to the entries in Table 1, our list of calibrating stars. These details also refer to the big list of short-period dwarf novae in Table 2, the main subject of this paper.

---

[2] And to some extent IUE. Large collections of WD temperatures, mainly derived from IUE spectra, are given by Winter & Sion (2003, Table 4) and Urban & Sion (2006, Table 2). Most are likely correct, and are a key contribution to the subject. But individually we give them somewhat lower weight, because many of the stars have a fairly high accretion luminosity, which makes the WD more difficult to distinguish from the inner accretion disk. We assign highest weight to cases where WD domination is certified by the spectrum, light curve, or (independently derived) $M_v$.

[3] Although boundary-layer emission spreading over most of the WD surface would be fine. The applicability of (1), and WD models in general, depends on the geometry and the temperature, not the true physical origin of the light.

### 3.1 Orbital Periods

Of the ~600 CVs with $P_{orb}$ now known, we judged that 46 meet the criteria for calibration: dwarf novae (magnetics need not apply) with sufficient constraints on $V_{max}$, distance, and binary inclination. Warner's study had 25 stars, of which we reject 10 as being insufficiently constrained. The remaining 15 are in both studies – although we used none of Warner's data or sources, in order to preserve independence.

### 3.2 Visual Magnitude at Maximum Light

In the 1980s, available data on dwarf-nova eruptions consisted of a blend of photographic and visual magnitudes. But we now have access to easily searchable variable-star records, especially that of the AAVSO. The human eye is the ideal detector for this purpose, since it is immune to changes in technology, and used by thousands of observers. Furthermore, the central wavelengths of the eye and the commonly used Johnson *V* filter are similar; and both detectors are broad enough to render line emission insignificant. Such records, together with the steady accumulation of published results, yield more accurate estimates of $V_{max}$, and for more stars.

In examining the visual data, mainly through the on-line AAVSO records, we averaged over 3–10 well-observed eruptions, after correcting a few errors and ignoring a few outlier points. Some of the latter are probably mistakes (e.g., from incorrect identification of the star). Some are known or surmised to be correct, but, in our opinion, unrepresentative because they are very brief excursions in a long eruption, or because they are "the brightest eruption ever". We would have liked to reduce this extra source of variance by estimating a *typical* $V_{max}$. This has extra pitfalls, however, because many dwarf novae show a bright/long and faint/short dichotomy, and the latter type is usually quite poorly observed. To reduce this extra variance, we chose to estimate $V_{max}$ for a "typical bright eruption". This usually meant special emphasis on the best-observed eruptions; in a few cases it also required a small correction for the (putative) unobserved peak of an eruption.[4]

We also used some data from previous large compilations (Glasby 1970; Patterson 1984, hereafter P84; Warner 1995; Ritter & Kolb 2003), with lower weight since we could usually not judge its consistency with the "typical bright eruption" standard.

---

4  With electronic notification now routine, observer vigilance tends to increase just after an eruption is announced. Most dwarf novae rise quickly to maximum light, so the peak is sometimes missed.

Column 3 of Table 1 contains the data. The first number gives $V_{max}$, and the second number gives $V_{max}$ corrected for extinction (all stars) and the effect of "supermaxima" (nearly, but not absolutely, all dwarf novae with $P_{orb}$ < 3 hours). In supermaxima, powerful superhumps suggest access to an energy source not available to non-superhumpers – likely tidal energy, extracted from the orbit. But it was not easy to ignore supermaxima and simply use normal maxima – because some stars have few or no normal maxima, and because supermaxima are in all cases *vastly better observed*. Based on 8 stars with extensive records of both, we judged that normal maxima are ~0.8 mag fainter than supermaxima.[5] So we measured $V_{max}$ from variable-star records ("typical bright eruption"), corrected for extinction, and added 0.8 mag to remove the estimated "extra light" of supermaxima.

### 3.3 Extinction

To estimate extinction, we used, wherever possible, E(*B–V*) estimates from interstellar absorption: either the ultraviolet continuum (the 2200 A "signature of graphite"), or the traditional atomic/molecular absorptions (Na I, 4430 A). Numerous E(*B-V*) estimates are compiled by Bruch & Engel (1994). But for these lightly absorbed stars, this information was usually absent or too crude. More commonly, we used infrared dust maps (Schlegel, Finkbeiner, & Davis 1998) for a through-the-Galaxy estimate, and then estimated the star's E(*B–V*) from a plausible assumed distance and an assumed model for dust distribution (Drimmel et al. 2003). We then assumed typical dust, with $A_v$ = 3.1 E(*B–V*). For most of the calibrating stars, $A_v$ did not exceed 0.2 mag. We used this to correct $V_{max}$ to the final values in column 3.

### 3.4 Distance

Distance estimates are given in column 4 of Table 1. These come primarily from the methods discussed in Section 2 above, applied to the data presented in the cited references. Our estimates sometimes differ from those in the literature, because we sometimes re-evaluate the authors' argument with today's improved calibrations (especially the K06 empirical mass-radius relation for the secondary). It is hard to assess the errors in these distances; except for the purely astrometric measures, they all depend on calibrations and astrophysical arguments. We estimate that typical errors are in the range 15-25%; asterisks indicate those with smaller errors *if* they are obtained from astrometry. See Section 3.6 below, and the catalogue description of Section 5, for a further discussion of errors.

Column 5 contains the corresponding $M_v$ at maximum light.

### 3.5 Binary Inclination

---

5  For CV fans *in extremis*, these are: YZ Cnc (11.3, 11.9); VW Hyi (8.7, 9.6); SU UMa (11.2, 12.0); Z Cha (11.8, 12.6); WX Hyi (11.5, 12.5); RZ LMi (14.4, 15.0); V503 Cyg (13.4, 14.0); and ER UMa (12.9, 13.6). The average difference is 0.77 mag, which we round to 0.8.

The brightness of an accretion disk depends strongly on viewing angle. For dwarf novae at or near maximum light, the radiating surface can be assumed to be optically thick, and in this case Paczynski & Schwarzenberg-Czerny (1980) found that for a binary inclination $i$, the angle-averaged magnitude of a flat, limb-darkened (with $u = 0.6$) disk should be corrected by

$$\Delta M_v(i) = -2.5 \log [(1 + 3/2 \cos i)(\cos i)]. \qquad (2)$$

This is a strong dependence: a pole-on disk is 1.0 mag brighter than an average disk at $i = 57°$, while stars with deep eclipses (typically $i > 80°$) should be at least 1.7 mag fainter than average. Therefore a correction is necessary.

This can be troublesome. Inclination estimates in the literature are often based on inappropriate (or inappropriately exact) assumptions: about the secondary's mass-radius relation, about the origin of the broad emission lines, about the origin of orbital signals in the light curve, about the origin of the periodic motions in radial-velocity. In general, observational clues permit only a coarse assignment of inclination: low, moderate, high, and very high – and only the two highest categories are really certain (because of eclipses). Even if observations yielded $i$ exactly, we still could not correct $M_v$ accurately for it; the opening angle and limb-darkening of CV disks are not well known, and each contributes an uncertainty not expressed by Eq. (2). Anyway, we did the best we could: surveyed the literature, accepted the well-supported estimates, and made our own estimates for others – based on the width and doubling of emission lines, plus the presence/absence of orbital photometric humps. Then we applied Eq. (2). The resultant estimates are given in columns 6 and 7 of Table 1, and the final corrected $M_v$ at maximum light is given in column 8.

### 3.6 Errors

We estimate the size and range of uncertainties as follows: 0.1-0.3 mag in $V_{max}$; 0.1-0.2 mag in $A_v$; and 10-30% in distance. These produce a combined uncertainty of 0.3-0.6 mag. What about the $\Delta M_v$ corrections for binary inclination? These vary from +0.95 to -2.6 mag – a mighty intimidating range! But the situation is not quite as scary as it looks. For small $i$, observations specify $i$ only roughly, but Eq. (2) is not sensitive to $i$; for large $i$, the sensitivity is great, but the light curves specify $i$ pretty well (from the eclipse waveform). Across all values of $i$, a realistic error on $\Delta M_v$ is probably near ±0.4 mag. The combined uncertainty for each entry in Table 1 is then 0.5-0.8 mag, depending mainly on the error in distance. While we do not "know the uncertainty" for each component element of each star, we have estimated the combined uncertainty in the corrected $(M_v)_{max}$ for each star in column 8. If realistic, this is sufficient for our purposes. Readers interested in the measures/errors of individual components should consult the primary works on these stars.

### 4. $(M_v)_{max}$ VERSUS $P_{orb}$

The upper frame of Figure 1 shows the resultant $M_v(P_{orb})$ relation:

$$(M_v)_{max} = 5.70 - 6.89\ P_{orb}\ (d)$$
$$= 5.70 - 0.287\ P_{orb}(hr), \quad (3)$$
$$(\pm 0.18)\ (\pm 0.018)$$

with an rms scatter of 0.47 mag. Despite the great improvements in data, and independence of both data and analysis, our relation is not appreciably different from that of Warner ($5.64 - 0.26\ P_{orb}$). The $M_v(P_{orb})$ relation therefore survives this more extensive test as a good distance estimator for erupting dwarf novae.

There remains some doubt as to whether addition of 0.8 mag to the magnitudes of supermaxima is appropriate. For these short-period stars, *most* of the accretion energy is released in superoutbursts, not normal outbursts; so maybe that +0.8 mag adjustment is unwarranted. Without the adjustment, a linear fit yields

$$(M_v)_{max} = 4.95 - 0.199\ P_{orb}(hr), \quad (4)$$
$$(\pm 0.16)\ (\pm 0.011)$$

with a similar rms scatter of 0.41 mag. This is shown in the lower frame of Figure 1 – and is probably more useful, since it refers to the actual measured magnitudes, not relying on the +0.8 mag correction.

These relations can be useful distance estimators for any erupting dwarf nova. But this paper mainly concerns those of short $P_{orb}$ (<3 hr), and inspection of Figure 1 shows that in this domain, $(M_v)_{max}$ of superoutbursts is practically constant at 4.60±0.14. This is a useful simplification. We can purify the sample further by considering only short-period stars with good distance measures (~15%); these are the stars with asterisks. Averaging these 9 stars and correcting for inclination, we obtain

$$(M_v)_{max} = 4.58 \pm 0.15. \quad (5)$$

Finally, since nearly all of these are SU UMa stars and such stars have well-defined "plateau" segments in their decline light curve, we can calculate and use the V magnitude on the plateau (defined in Section 9 below, and presented in Table 2) as a distance estimator. Measurement of 18 short-period calibrators[6] in Table 1, together with corrections for distance and inclination, yields

$$(M_v)_{plat} = 5.5 \pm 0.2. \quad (6)$$

The plateau can sometimes be a better constraint, because the brightness is high and long-lasting, whereas the $V_{max}$ peak is sharp and sometimes unobserved. In practice, our "distance by $M_v$" estimate, which figures extensively in Table 2, uses all three constraints [Eqs. (4), (5), and (6)], weighted as the data quality warrants. The tight clustering of these estimators, and in Figure 1, shows that the dwarf-nova outburst is really a very uniform phenomenon.

---

6  Inadequate observation, or the failure to show clear superoutbursts, disqualifies 4 of the 22 stars: IR Com, V893 Sco, BZ UMa, and SDSS J1227+51.

The evidence supports that uniformity for the 46 stars. To the extent that all dwarf novae agree with it, the corresponding estimates for the distance $d$ (in parsecs) are

$$\log d = (V_{\max} - A_v + 0.4)/5 \qquad (7)$$
$$\log d = (V_{\text{plat}} - A_v - 0.5)/5. \qquad (8)$$

It should be remembered that these apply to stars with an average binary inclination ($i=57^0$). For the least-studied stars, no correction is possible. When $i$ can be estimated, correction via Eq. (2) is desirable. But more commonly, even when no real estimate of inclination is feasible, we always know whether or not a particular binary is deeply eclipsing. Default estimates of $i= 80^0$ ($\Delta M_v = -1.7$) and $i=49^0$ ($\Delta M_v = +0.3$) are then recommended, respectively. And there is one more caveat. These standard candles were derived from studying ~200 outbursts (averaging 4-5 for each of 46 stars). For any one outburst, and especially for outbursts which may not have been caught at maximum light, less weight should be placed on them.

## 5. PROPER MOTIONS

The earliest discussions of CV distances (e.g. Kraft & Luyten 1965) used statistical parallaxes (i.e., radial velocities and proper motions) to establish mean absolute visual magnitudes for dwarf novae: +4.5 in eruption, and +7.5 in quiescence. These estimates became very famous, and indeed polluted some of the later discussion about distance, when authors uncritically adopted these estimates. By the mid-1980s, it was clear that some dwarf novae are much fainter in quiescence, because the spectra and/or light curves revealed WDs of modest temperature (~15000 K). Strangely, recent authors have generally paid little attention to proper motion, even though there are many accurate measurements available, and there is distance information hiding in the proper motions.

Since distances are precious, we try here to make some use of that proper-motion data. Of course, the principle is that rapidly moving stars are probably nearby. The Dartmouth astrometry program yields reliable measures for a few hundred stars (T08, Peters 2008); the merged NOMAD set of various automated USNO surveys (Monet et al. 2003, Zacharias et al. 2004) yields even more stars, though somewhat at the expense of errors and accuracy.

We make only a primitive use of proper motions: comparing them to the distance estimates of Table 1, and the better-quality distance estimates of table 2 (described below in §7).[7] This comparison is shown in Figure 2, which includes CVs other than dwarf novae (magnetic variables, classical novae, novalikes) since these additional stars probably arise from a similar underlying population. It also includes stars with $P_{orb}$ > 3 hours (from a separate catalogue not discussed in this paper). As expected, proper motion scales inversely with distance, and the slope of the line for all stars suggests $v_{tang}$ = 44±5 km/s. This could be considered a tertiary distance clue.[8] Its merit is that it applies to hundreds of stars and is purely astrometric. Its demerits are: (1) it is purely statistical, not furnishing a strong constraint on any individual star; (2) it effectively only constrains stars of high proper motion; and (3) it assumes that all CVs are drawn from the same population, which is unproven, and known to be false in detail.[9]

After finishing this paper and identifying ~20 stars as candidate "period-bouncers" in Table 3, we returned to the proper motions and considered these 20 separately, along with 2 others which are good candidates but disqualified because of their obvious magnetism (EF Eri and EX Hya). These are shown as the "B" (for bouncer) symbols in Figure 2. The marked line shows the fit for the 121 other stars, and implies $v_{tang}$ = 39±5 km/s. The bouncers suggest a line of somewhat greater slope, more like 57 km/s (depending on how the nearest star, WZ Sge, is handled).

If true, this might indicate that period-bouncers represent a somewhat older population, in accordance with the age-velocity relation of stars in the solar neighborhood (see Wielen 1977, Kolb & Stehle 1996). We will see other evidence of this below, in the stars' distribution in the Galaxy (§ 15).

## 6. OTHER DISTANCE CLUES

Other distance clues have been occasionally used in previous studies:

(1) the expansion parallax of classical nova shells;
(2) the standard candle of classical novae in full eruption;
(3) the relationship between $M_v$ and the equivalent width of disk emission lines;
(4) interstellar reddening/extinction; and
(5) location in the Galaxy.

---

7  Some of these are "primary" - meeting the standards of Section 2, but missing other requirements for inclusion in Table 1: not dwarf novae, no previous eruption, or insufficient data concerning $V_{max}$ and inclination. Others are secondary: with excellent DN data, but a distance relying on the $M_v(P_{orb})$ relations.
8  For this study we used distance estimates which do not include proper motion – viz., from the three primary techniques, plus the standard candle of the DN eruption, when $V_{max}$ (or $V_{plat}$) and inclination are adequately specified by observation. The final tabulated distance estimates are slightly different, because they include some low-weight contribution from the several tertiary methods discussed in this paper. A more sophisticated use of proper-motion data, in conjunction with trig parallax, is discussed by T03.
9  In particular, among H-rich CVs there are two obvious outliers, SDSS J1507+52 and BF Eri (Patterson et al. 2008, Sheets et al. 2007). These have been excluded since they show very high values of $v_{tang}$.

These clues are much less useful for our purposes. (1) and (2) are useful for classical novae, but we exclude these from consideration; our emphasis is dwarf novae, and recent classical novae are often too agitated by the eruption to have reached their pre-eruption state. (3), (4), and (5) supply only very rough clues, pertinent mainly "when all else fails". We did not use any of these clues in calibrating $M_v(P_{orb})$; but where other constraints are coarse, we used (3)-(5), with low weight, in estimating distance for the larger sample of stars in Table 2 (below).

Since the equivalent-width relation has not been updated in many years, we show a modern calibration in Figure 3. The rough lesson of its original version (Figure 6 of P84) remains: the equivalent width of hydrogen emission lines increases as the accretion disk grows more intrinsically faint. This describes every *individual* star's variation through the eruption cycle, and star-to-star variations as well. The explanation is probably simple: that the low-density disks of quiescence are mostly optically thin, and line emission therefore contributes significantly to their cooling (Tylenda 1981). This can be a helpful clue in obtaining a rough distance estimate. It is, however, confounded by binary inclination, which is likely to affect both $M_v$ and the strength of line emission, and likely in opposite ways. High inclination definitely suppresses continuum, roughly according to Eq. (2). And eclipsing CVs teach us that line emission comes predominantly from regions well above the disk plane, because line fluxes tend to remain relatively unchanged across eclipse. So very high *i* implies that we'll see all the line flux (maybe *double*, because we can see below as well as above the disk) and no continuum – even if the star's intrinsic continuum greatly outshines its line emission (as measured by imaginary observers scattered over 4π sr). Therefore inclination, as well as the low density and low $\dot{M}$ characteristic of dwarf-nova quiescence, can move stars significantly to the upper right of Figure 3. Given this sensitivity and the large scatter, equivalent width is best regarded as a *rough* clue to $M_v$.

## 7. A CV CATALOGUE

For the past decade we have been preparing a catalogue of all CVs of known $P_{orb}$. The rapid increases in data and membership have made it hard to complete! But a working portion of it, with observed and inferred data relevant to this paper, and confined to known or likely dwarf novae with $P_{orb}$ < 3 hours, is available at http://cbastro.org/dwarfnovashort/. We shall call this Table 2, and the first few entries are shown in the printed version of this paper. As of 6 October 2010, the online table contained 292 stars.[10] We know less about stars which don't meet the elite standards of Table 1; but there are a lot more of them, and they supply many useful clues. The remainder of this paper relies on data from Table 2 – and especially data relevant to *distance*.

We systematically exclude magnetic CVs: the AM Hers (polars) and DQ Hers

---

10 Included also are a few stars with $P_{orb}$ still unknown, but very likely relevant to this collection – probably with short $P_{orb}$, and possibly some period-bouncer credentials.

(intermediate polars).  The physics of accretion in these stars is very different, and the methods of discovery and study are very different, with much of the luminosity radiated in an EUV/soft X-ray component which is easily blocked by the interstellar medium – or even by circumstellar material.  Dwarf novae are much tamer, with a standard candle, a regular morphology of high/low states ("eruptions"), and a large army of visual observers who have tracked these stars for many decades.

   Table 2 contains the following data/estimates:

Column 1.  The GCVS name of the star, any commonly used alternate name (or a shortened version), and approximate equatorial coordinates (*hhmm+dd*).  When no GCVS name is yet assigned, the star is usually labelled for the survey which revealed it, according to the following abbreviations:

SDSS = Sloan Digital Sky Survey (York et al. 2002)
RX = Rosat All-Sky Survey (Voges et al. 1999)
ASAS = All-Sky Automated Survey (Pojmanski et al. 2003)
CSS = Catalina Sky Survey (Drake et al. 2009)
OT = optical transient (found in miscellaneous variability searches, usually by amateur
      visual/photographic observers).

Column 2.  The first row is the max-min range of the *V* magnitude.  With no further information, this is assumed to arise from accretion light.  Where photometric and spectroscopic information warrants, we subtract the estimated non-accretion component (usually from the WD), and the second row gives the accretion component of the quiescent *V.*  This point is discussed further in § 8 below.  Where applicable, an "E" indicates a deep eclipse; this is important since high inclination diminishes the flux received from a flat disk.  (But all listed magnitudes are out-of-eclipse).

Column 3. The first row gives the orbital period in days, with a letter code indicating its source. In order of decreasing reliability:
 e = based on eclipses (probably 100% reliable);
 v = based on radial velocities (very reliable);
 p = based on a stable photometric wave at quiescence (usually reliable)[11];
 o = based on an outburst orbital hump (also sometimes called "orbital superhump" or
     "early superhump"; known to be valid in a few cases, but the universality and
     accuracy are not known);
 h = based on sidebands in the superhump power spectrum ($n\omega$-$m\Omega$, where *m* and *n*
     are unequal integers; known to be valid in a few cases, but the universality and
     accuracy are not known).

---

11 A truly stable period at true quiescence should nearly always be $P_{orb}$.  Sometimes the observational baseline is too short to prove high stability.  Sometimes, especially for the ER UMa stars, there is a worry that the photometric signal is a long-lived residue of a superhump from the preceding eruption – because in a few cases (WZ Sge and GW Lib), the superhump is known to last hundreds of days. This may have led to an erroneous $P_{orb}$ reported for KV And (Patterson et al. 2003, hereafter P03), because a second visit to that star showed no photometric wave.

The second row gives the superhump period[12] in days. The third row gives the rough Galactic coordinates (longitude + latitude).

Column 4. The first row gives the variable-star type. Most are SU UMa-type dwarf novae (with supermaxima accompanied by superhumps). Others are labelled as follows:

ER = ER UMa stars (a rare subclass of SU UMas, with very frequent eruptions and little or no time at quiescence).
WZ = WZ Sge stars (a rare subclass of SU UMas, with very infrequent supermaxima).
DN = stars showing dwarf-nova outbursts, but not yet clear supermaxima.
NL = "novalike" stars, remaining in a low state but likely to have a future supermaximum.
GW = GW Lib stars (WZ, DN, or NL, with white-dwarf pulsations).
CN = classical novae, with year of eruption.
UX = UX UMa stars, which stay in a bright state. (We use this term to indicate "bright state" rather than the more restrictive traditional meaning, which requires broad accretion-disk absorption lines.)

The second row gives the superoutburst recurrence period in days. This important number is mainly drawn from inspection of variable-star archives (AAVSO, VSNET, BAA, RASNZ, ASAS, CSS), with some weight also given to previous tabulations. The 3-5 month gap between observing seasons is a big problem, so the estimate is sometimes quite uncertain. This point is discussed further in § 9. The third row gives the estimated binary inclination in degrees, from the criteria discussed in – and subject to the worries of – § 3.5.

Column 5. The first row gives the estimated mass ratio $q = M_2/M_1$, usually obtained from the superhump period excess $\varepsilon = (P_{sh}-P_{orb})/P_{orb}$ and an adopted $\varepsilon(q)$ calibration (P05). The accuracy of the calibration is least reliable at low $\varepsilon$. Some, marked by *e*, come from eclipses; these are probably more accurate, though subject to the assumptions used in analyzing the eclipse geometry. Since the $\varepsilon(q)$ relation is based on eclipsers, these measures of *q* should be compatible. Some, marked by *v*, come from radial-velocity measurements. The third row gives the equivalent width of Hβ emission,

---

12 Superhump periods change slightly, so what is meant by "the" superhump period? In the past I have used the criterion *estimated period 4 days after maximum light* (*or hump onset*), for several reasons:
   (1) It's a time of large hump amplitude and good observer diligence, so $P_{sh}$ tends to be well-defined.
   (2) It's a time when $P_{sh}$ is not changing rapidly.
   (3) It's a time (plateau phase) when most of the eruption energy is radiated.
   Admittedly, the O-C curves show details going far beyond this simple characterization (Kato et al. 2009, hereafter K09). But this is our best effort at "one number per star". In the language of K09, these would be called "stage B superhumps".

   In addition, there are slight variations in $P_{sh}$ between superoutbursts – partly intrinsic and partly from the accidents of observational coverage. Entries in column (3) are basically a weighted average, from published data and the CBA archives. We state four significant figures, because the uncertainty is always near 0.0001 d.

in the star's normal luminosity state (usually quiescence).

Column 6. The first and second rows give the proper motion in mas/year, from the sources discussed in § 5. Errors are typically in the range 5-10 mas/yr. The third row gives the γ-velocity (in km/s) from radial-velocity studies.

Column 7. The first row gives the estimated distance and error, from the totality of evidence (trigonometric parallax, WD $M_v$, WD fit to spectrum, Bailey relation, DN standard candle, proper motion, etc.) discussed in § 2-6. Errors below 15% generally arise from a good-quality parallax; in this case, the distance and error are usually verbatim from the cited work. Distance estimates with errors in the 15-25% range usually come from a medley of constraints, sifted by the author – and are stated in round numbers to emphasize that these are human judgments. Slightly larger errors (25-35%) are estimated in cases where at least two of the important and common clues ($V_{max}$, $V_{plat}$, $V_{WD}$, and $i$) are poorly constrained. The second row gives the estimated absorption $A_v$ along the line of sight (see § 3.3). The third row shows the clues utilized for each star, according to the following code:
 1 = trigonometric parallax
 2 = Bailey method
 3 = WD photometric or spectrophotometric parallax
 4 = $(M_v)_{max}$-$P_{orb}$ relation
 5 = proper motion
 6 = equivalent width ($M_v$)
 7 = classical-nova properties.

Column 8. As discussed in § 9, the first row is the $V$ magnitude of the "square-wave" equivalent of the plateau segment of an average superoutburst. The second row is the duration of that square wave in days. We estimate from detailed accounting of the best-observed light curves that the total $V$ light radiated in the eruption is ~30% greater than the $V$ light radiated in this interval.

Column 9. The first and second rows give the FUV and NUV magnitudes, on an "AB" magnitude scale. Most of this data comes from the GALEX archive; some is synthesized from HST/IUE spectra. The third row gives the WD temperature, deduced from a fit of the ultraviolet spectrum to a WD model atmosphere.

Column 10. The first row gives the estimated $M_v$ of accretion light at quiescence. (However, the WD component of quiescent light has only been removed when it is measurable in the spectrum or light curve; so other entries implicitly contain an unwelcome and unsubtracted WD component.) The second row gives <$M_{ve}$>, which corresponds to the time-averaged flux of eruption light (including the "0.3 mag" correction discussed below in §9, but *not* including any quiescent flux). The third row corrects that time-averaged $M_{ve}$ to a standard binary inclination of 57°.

Column 11. References and notes. These are the sources most directly used in preparing this table, not necessarily the most complete or up-to-date reference. Common abbreviations are K09 (Kato et al. 2009, a large collection of superhump studies) and CBA (unpublished Center for Backyard Astrophysics data).

## 8. QUIESCENT $M_v$ VERSUS $P_{orb}$

Nearly every dwarf nova has a catalogued $V_{max}$, and about 400 have a known $P_{orb}$. Eqs. (3)-(6) therefore furnish a rough distance clue for ~400 dwarf novae, along with the other detailed clues discussed in § 2–7. We can then estimate $M_v$ at quiescence, and study its dependence on $P_{orb}$, for hundreds of stars.

### 8.1 Corrections for Non-Accretion Light

Since disks brighten to $M_v$=+4.6 during eruption, they then outshine any sources of nonaccretion light. This is why our measure of eruption light ignores contributions from the secondary star and WD. But in quiescence these corrections can be important, and difficult to estimate. We give it our best shot in column 2 of Table 2. The catalogue's $P_{orb}$ upper limit of 3 hours eliminates the need to correct for the secondary's brightness, since eclipse light curves demonstrate that the secondary contributes only a few percent to the V light of short-period dwarf novae. But the WD can be an important contributor to the quiescent light. The signatures are well known: sharp eclipses, broad Balmer/Lyman absorptions in the spectrum, and sometimes nonradial pulsations in the light curve. In the simplest parsing of CV light, the WD component is a non-accretion source.[13] Therefore we examined available spectra and light curves of all stars, estimated the WD contribution, and then subtracted that (if measurable) from the observed quiescent V flux. We made no subtraction for stars lacking any obvious WD signature – although improved spectra would certainly reveal the need for that subtraction.

### 8.2 Exclusions and Inclusions

Although every nonmagnetic CV with $P_{orb}$ < 3 hours is tabulated (since they are all good dwarf-nova candidates), several classes of CV need to be excluded in this study of quiescence. For at least a few decades after eruption, the remnants of classical novae often show high-excitation spectra and an anomalously bright $M_v$ – suggesting access to a more powerful energy source. This may be light arising from the reprocessing of UV radiation from the WD, rendered hot by the recent thermonuclear runaway. We rejected these stars altogether, as not reliably representative of pure accretion light.

---

13 Although this is not correct in detail. We now understand that WD temperatures are raised by compression and accretion heating, and these depend on $P_{orb}$ because accretion rate does (Sion 1999, Townsley & Bildsten 2003). Complicating this further is the discovery that the WD cools significantly between outbursts, even on timescales as long as decades (Sparks et al. 1993, Slevinsky et al. 1999, Godon et al. 2006).

In considering the quiescent magnitudes, we also excluded a few stars which are otherwise honorable dwarf novae, but which "never reach quiescence".  These are usually called the ER UMa class (Robertson, Honeycutt, & Turner 1995).  They show rounded light curves, or stay only a few days, at minimum.  In some cases, the apsidal superhump persists through their brief stays at minimum[14]; this periodic feature is known to be a signature of eruption, and may arise from an extra energy source (tidal) in the disk.  This behavior disqualifies the ER UMas in a study of dwarf-nova quiescence, and implies that their periodic photometric waves at minimum might be merely a residue of the superhump, not accurately signifying $P_{orb}$.  We also excluded the few "UX UMa stars" in the table; these stars resemble dwarf novae in permanent and full eruption, and therefore pollute a study of quiescent magnitudes.

On the other hand, we included – in the table and in the analysis presented below – stars with no recorded eruption, if their properties appear to be otherwise similar to quiescent dwarf novae (nonmagnetic, with broad H emission lines suggestive of accretion through a disk).   These stars have no known value of $V_{max}$, but they are useful in studying the quiescent V light, since they are excellent candidates for a future eruption, and many have good-quality distances arising from the measurable presence of the WD.

### 8.3 Quiescent $M_v$

In Table 2, row 1 of column 2 gives the max-min range of the star, and row 2 gives the quiescent V magnitude $M_{vq}$ after subtracting, if appropriate, the estimated V flux of the WD. These entries are drawn from the 50-year published record of calibrated V magnitudes, from the 100-year record of visual observers, and from our own unpublished data.  In column 7, row 1 gives the distance and error, estimated mainly from the methods listed in row 2, and row 3 gives the estimated $A_v$ (with both quantities iterated once).  The top row of column 10 then gives the estimated accretion component of the quiescent $M_v$ (although we emphasize again that some fraction of the WD flux is probably accretion light leaking out on long timescales).

No correction for inclination has been applied to the quiescent magnitudes, however.  The quiescent disks are emission-line sources, and therefore must be very different in geometry and optical depth.  For the accretion rates relevant to quiescence, the disks are almost certainly optically thin in the continuum (Tylenda 1981), so Eq. (2) is inappropriate.  Inclination corrections can be added to the long list of unanswered questions concerning quiescent disks.[15]

---

14 This is known to be true for one star, RZ LMi.  Whether it is true for the others is an important unsettled question.  See also footnote 11.  Between supermaxima, two likely members of this class – IX Dra and DI UMa – show apparently stable photometric waves, which would imply very low values of $q$ if they signified $P_{orb}$, *and* if such stars obey the standard $\varepsilon(q)$ relation.  A radial-velocity study for IX Dra would clarify this.  Anyway, we exclude these stars, since they apparently never reach quiescence.

15 Although we did not correct for inclination, we did exclude deep eclipsers.  We found that quiescent accretion light in those stars was much fainter than that of their non-eclipsing comrades – probably

8.4 Quiescent $M_v$ Versus $P_{orb}$

The correlation of $M_{vq}$ with $P_{orb}$ is shown in Figure 4. Where $P_{orb}$ is not known precisely, we estimate it from $P_{sh}$, using the ε(q) and $q(P_{orb})$ relations of P05. (Typically this means a $P_{orb}$ estimate good to 1-2%, and this error is unimportant for all subsequent purposes in this paper.) Stars within 100 pc are given a bigger symbol. Why? Because we shall find later, mainly in §15, that distant CVs are greatly undercounted, and the solar-neighborhood residents are more likely to represent the true (unknowable) population. Thus we give them more visual prominence, to illustrate how the distribution might change if the census were complete.

The only obvious trend is that stars of shorter $P_{orb}$ tend to be fainter. In addition, the 100 pc sample suggests that a complete census would have a much higher percentage of intrinsically faint stars. The plot's usefulness is limited by the dependence of $M_v$ on poorly known quantities: distance, the likely and uncorrected effect of binary inclination, and simple measurement error of the quiescent $V$ (which unfortunately often depends on one snapshot magnitude, at an unknown time in the eruption cycle). The origin and amount of scatter is discussed further in § 11.

With considerable mystery about the source of accretion light in quiescence, it's hard to say what $M_{vq}$ really represents physically. But all the stars are likely dwarf novae, the great majority are proven dwarf novae, and dwarf novae appear to be a very uniform class (differing only in their recurrence times). So it's reasonable to hope that the detailed physics which allots quiescent accretion luminosity to $M_v$ might be fairly similar from star to star – which would allow $M_{vq}$ to be a qualitative surrogate for luminosity and accretion rate. It would be nice if future detailed study could evaluate this hope, or even make it a *quantitative* surrogate.

## 9. <$M_v$> FROM ERUPTION LIGHT

The mysteries at quiescence presently put a question mark on inferences based on $M_{vq}$. But there is also information from eruption, where data is plentiful and the basic physics is pretty well-known: release of gravitational energy in a viscous disk. Nearly all the successes[16] of accretion-disk theory come from the behavior of dwarf novae in eruption. And most of the light does too: studies of the best-observed dwarf novae show that most of the light is emitted in outburst (Osaki 1974, Smak 1976, Smak 1984, P84, Anderson 1988). Therefore a measure of the total light (or better yet, bolometric luminosity) would be very useful. In the P84 study, we integrated under each star's

---

because an inclination correction is needed (for all stars), and/or because much of the disk is actually obscured by structures in the outer disk (Knigge et al. 2003). We excluded these 17 stars.

16 Except for the first success: the double-peaked emission lines, which loudly broadcast the presence of a disk. After that realization in the 1950s-60s, new discoveries about the accretion structures in quiescence have been mostly unanticipated, and often puzzling.

historical visual light curve, applied a suitable bolometric correction, corrected for distance, and thereby obtained an estimate of <$L_{bol}$>. This worked pretty well, and revealed a strong correlation of $L_{bol}$ – and therefore $\dot{M}$ – with $P_{orb}$ (Figure 7 of P84). But for short-period dwarf novae, the quality of this estimate was poor: few distance estimates were then available, and only sparse data on the duration and recurrence times of eruptions.

The situation is much better today, and we have repeated the P84 study in a simplified form, with data in columns 4 and 8 of Table 2. In doing this study, we found that most of the visual light of short-period dwarf novae is emitted in the "plateau" phase of superoutbursts. There are some variations – between stars, and between the superoutbursts of a given star – but the following pattern is nearly universal. The star rises quickly to a repeatable maximum $V$ (corresponding to $M_v$ = +4.6), quickly declines ~0.3 mag, settles to a slow linear decline at a rate of 0.12 mag/day, and finally falls quickly after 10–25 days. The plateau is the slow and roughly linear decline. Column 4 contains the star's classification and recurrence time (for supermaxima) in days, gleaned from the historical records of the AAVSO and other visual observers. The plateau phase is pretty well-defined, subject mainly to uncertainty about precisely when the eruption began. Since the plateau phase supplies most of the eruptive flux, has a repeatable shape, and is well observed, we have used that data from historical records (and from papers in the scientific literature when an eruption is sufficiently well covered). More precisely, we approximated each eruption as the flux equivalent of a square wave of height $V_e$ and duration $T$. These estimates are given in column 8 of Table 2. We then divided by the recurrence period $T_{rec}$ to obtain the time-averaged $V$ flux of eruption light, and the corresponding magnitude <$V_e$>. For example, a square wave at $V$ = 10, lasting for 12 days, yields <$V_e$> = 15 if $T_{rec}$ = 1200 days. In the best-studied cases, restriction to the plateau phase seems to underestimate the total <$V_e$> by 0.2–0.5 mag (because of rising light, echo outbursts, normal outbursts, and the long, fading tail of superoutbursts). The physics of these phenomena is diverse and not necessarily understood, but the energy source is likely accretion, so a correction is required. For simplicity we applied a uniform 0.3 mag correction for "other" accretion $V$ light. Then we used a distance estimate to calculate <$M_{ve}$>. This is contained in row 2 of column 10. For stars with an inclination estimate, we then used Eq. (2) to convert to a value at $i$ = 57°; this is in row 3 of column 9.

Stars erupting only once present a challenge, because we need to estimate a lower limit to $T_{rec}$. This requires close attention to the various sources of eruption data, mostly through web/email archives: AAVSO, ASAS, VSNET, BAA, RASNZ, CSS, and the 20–40 visual and CCD observers worldwide who regularly monitor these very inactive stars. We estimate a lower limit on that basis. A few stars have never erupted, and yet have been observed long enough (>5–10 years) to yield a useful lower limit to $T_{rec}$ (equal to about half the observation time for a frequently-monitored star, but less stringent if the coverage is sparse).[17] We also need to adopt an estimate of the

---

17 More precisely, we assume that observing records with gaps <10 d will catch all eruptions, and that (imaginary) visual light during a plateau phase. To do so, we averaged the inclination-corrected values found for six well-studied WZ Sge stars: WZ Sge, GW Lib, V455 And, VY Aqr, AL Com, and EG Cnc. On average, these superoutbursts rise to $M_v$ = +4.5±0.3, and the plateaus (at 5.5±0.2) last for 19 days. The plateau flux is then diluted by the factor (19 days)/$T_{rec}$. Finally the star receives credit for an extra 0.3 mag from other accretion light. From these assumptions and a distance estimate, a few useful limits can be obtained for the nonerupters, and these are also given in row 3 of column 10. No inclination correction is needed, since the plateau at $M_v$ = +5.5 is already computed for a binary inclination of 57°. These considerations imply an estimate of

$$<M_{ve}> = 2.1 + 2.5 \log T_{rec} \text{ (d)}. \tag{9}$$

The resultant plot of corrected $<M_{ve}>$ versus $P_{orb}$ is shown in Figure 5. Stars of shorter $P_{orb}$ are much fainter, as in Figure 4; and the sprinkling of very faint stars, even at fairly long $P_{orb}$, gives the distribution some resemblance to the "boomerang" shape of $q(P_{orb})$ (e.g. Figure 9 of PTK).

## 10. Q VERSUS $P_{orb}$

Since the secondary star loses mass monotonically in the course of evolution, a plot of $M_2$ versus $P_{orb}$ could have great significance, and could show the arrow of evolution. As discussed many times previously (Patterson 2001, hereafter P01; P03, P05, PTK), the mass ratio $q$ is an excellent surrogate for $M_2$, because $M_1$ does not stray too far from 0.8 $M_o$, and because $q$ is the quantity more directly, accurately, and frequently yielded by dynamical studies and their close cousins (eclipses, radial velocities, superhumps). The observed $q(P_{orb})$ curve has provided a good constraint on evolution, and even an accurate mass-radius law for secondary stars in CVs (P05, K06). Figure 6 presents an updated version. The dots are values of $q$ inferred from superhumps (126) or eclipses (15), and the triangles are upper limits inferred from radial-velocity curves according to the discussion/prescription of PTK. This is a much cleaner representation of period bounce (at $P_{orb}$ = 0.054 d, with possibly a ±0.002 d intrinsic spread), and of a lower branch of CV evolution (containing 15-20 points and 4-5 upper limits).

The superimposed curve shows the predicted trend for a popular assumption: that evolution is driven by angular-momentum loss equal to that radiated by gravitational waves ($\dot{J}_{GR}$), with a secondary assumed to have a main-sequence structure at long $P_{orb}$, but then suffering increasingly from thermal imbalance towards shorter periods (e.g., Kolb & Baraffe 1999). As discussed many times previously (Patterson 1998, P01,

---

eruptions occur in the off-season with the same probability as during the observing season. A season typically lasts 8 months, gaps reduce it to 6 months, and thus a 10-year record is roughly equivalent to 5 years without gaps. However, the yearly spacing of gaps insures that a few large errors in the $T_{rec}$ estimate will occur, when the seasonal timing of eruptions has particularly bad (or good) luck. In the early days of DN studies, there were an amazing number of $T_{rec}$ estimates near 365 days.

P03, PTK, L08, Barker & Kolb 2003), this assumption yields a value of $P_{orb}$ at period-bounce which is too low, and values of $q$ which are too high, compared to the measured values. If the points in Figure 6 actually represent a boomerang-shaped *evolutionary* curve, then bounce seems to occur in the range $q = 0.065$-$0.080$. Adopting an average WD mass of 0.75-0.85 $M_o$ (K06, L08), this corresponds to $M_{bounce}$=0.049-0.068 $M_o$.

## 11. *Q* VERSUS <*$M_{ve}$*>, AND FIGURE 6 REVISITED

So we have one good delineation of a boomerang shape, and one which could at least be interpreted that way (<$M_{ve}$>). Uncertainties in distance and binary inclination presumably muddied the waters for Figures 4 and 5, whereas $q$ is independent of distance and $i$, and is more or less an observable quantity, whenever a star shows well-defined superhumps.[18] But if these figures are to be considered additive evidence for period bounce, then we should also require that the lower-branch candidates must be **the same stars**. If they are, then <$M_{ve}$> should correlate with $q$, at least for low $q$. Does it?

Well, yes. The dots in Figure 7 represent positive measurements from Table 2. The triangles show upper limits on <$M_{ve}$>, usually because two consecutive superoutbursts have not been seen. The dots with arrows show upper limits on both <$M_{ve}$> and $q$. Again we restrict consideration to normal dwarf novae, excluding the ER UMa and UX UMa stars, and old novae. A well-defined correlation appears – establishing that the same stars are producing the general trends in Figures 5 and 6. All the stars of very low $q$ are also very faint – as required. The general trend is shown by the line, which corresponds to $L_v \sim q^{2.4\pm0.2}$.

The good fit in Figure 7 is sufficiently encouraging to warrant calculating a surrogate $q$, inferred from the Figure 7 trend, for stars with a measured <$M_{ve}$>. This adds 41 stars to Figure 6, and we show an amended version in Figure 8, where open symbols indicate values of $q$ calculated from <$M_{ve}$>, assuming the stars are "normal". Since none of the 108 stars in Figure 7 appear "abnormal" (the rms scatter is just 0.9 mag), that does not seem too outrageous an assumption. Still, these 41 estimates are obviously of lower quality and independence, since they depend on distance, not dynamics, and are constrained to lie on the trend line of Figure 7.

Figure 7 is also relevant to an important question not otherwise considered in this paper: are there large long-term (millennia?) swings in a CV's accretion rate? That could wreak havoc on attempts – like the present paper – to constrain evolution by brief snapshots (decades) of a CV's observational record. But it would also cause a wide spread in <$M_{ve}$> at every $q$, apparently contrary to Figure 7. To some degree, this admittedly arises from excluding the ER UMa/UX UMa stars, which have a bright $<M_{ve}>$ even at low $q$. But such stars are extremely rare – only 12 of the 270 stars in Table 2, and likely <1% of the population in space density (because their superior luminosity selects them for discovery at greater distances; see § 14 for discussion). So for the vast majority of a CV's lifetime, it seems likely that $q$ determines the star's luminosity state.[19]

---

18 Assuming an $\varepsilon(q)$ relation. We used Eq. (8) of P05, an empirical fit from observations of eclipsing binaries. Other relations have been proposed, with some additional theoretical support (Pearson 2006, Smith et al. 2007). The differences are too small to affect our results, but are certainly worth exploring at very low $q$.

Since Figures 6-8 look quite good – consistent with the idea that CVs move along a well-defined evolutionary track – the ugly cosmetics of Figures 4 and 5 deserve a comment. Both $M_{vq}$ and $<M_{ve}>$ depend on distance and binary inclination, and we have estimated that these contribute an uncertainty of 0.7 mag even in fairly well-studied systems (Table 1). In the proletariat of Table 2, the standard-candle estimator [$(M_v)_{max}$ = 4.6±0.2 and $(M_v)_{plat}$ = 5.5±0.3 (corrected for absorption and inclination)] is additionally available, but the observational data on each star ($V_{max}$, $V_{plat}$, $V_{qui}$, $T_{rec}$, and especially $i$) is usually much sparser. We estimate a typical combined scatter of 0.9 mag from the uncertainties in these ingredients. For $P_{orb}$>0.065 d, the measured rms dispersion [about a smooth $M_v(P_{orb})$ curve] is 1.4 mag in $M_{vq}$ and 1.0 mag in $<M_{ve}>$. This suggests that the basic data of the light curve, along with distance and $i$, are probably the dominant sources of uncertainty – with some extra unknown factor contributing to scatter in $M_{vq}$ (possibly inclination, since we did not dare to correct for it).

So that partly explains the ugliness of Figures 4 and 5. For $P_{orb}$<0.065 d, the observed dispersion is much higher: 1.7 mag in $M_{vq}$, and 1.4 mag in $<M_{ve}>$. These are basically equivalent to *no correlation at all*. Since Figure 7 proves that $L_v$ varies strongly with $q$, and $q$ generally varies with $P_{orb}$, why should that be? A plausible answer could be: **because there are period-bouncers**. With the short-period regime populated by stars which are coming and going, the preferred diagnostic would be a measurable quantity which varies monotonically with age and comes from dynamics – independent of distance, inclination, and magnitude.

That would be $q$. It's hard to learn $q$ – but worth it!

## 12. WHITE DWARF TEMPERATURE VERSUS $P_{orb}$

Nearly every gram which moves inward through the disk should land on the WD, thereby heating it through compression and accretion. In theory, a correct account of this heating can predict WD temperature as a function of accretion rate; and by comparing the observed spectra to WD model atmospheres, one can learn $T_{WD}$ and therefore $\dot{M}$. This has several advantages over the use of $M_v$. It is independent of distance and *i*, and tends to smooth over the hurly-burly of eruptions, because the heat reservoir of the WD vastly exceeds that of the disk. There are theoretical advantages too, because $\dot{M}$, not $M_v$, is the ideal reward: the quantity which controls the rate and direction of evolution. Many efforts have been made to exploit the observational estimates of $T_{WD}$ from ultraviolet spectra (Sion 1999, Townsley & Bildsten 2003, Townsley & Gansicke 2009).

---

19 Although this mitigates the concern about mass-transfer swings, it does not cancel it. Even 1% of a star's lifetime is significant if the accretion rate is then 100x higher; that would effectively double the true average accretion rate. Our estimates of the ER UMa and UX UMa luminosities are bright enough ($<M_{ve}>$= +7.2 and +5.2 respectively) to make this a concern.

In the fullness of time, this measurement – relying on ultraviolet spectra – probably will become the best way to learn the accretion rates. However, it is now 33 years since the launch of IUE, and 22 since the launch of HST; and the most recent effort to compile and interpret the data (Figures 5 and 6 of Townsley & Gansicke 2009) is still ambiguous, due to scarcity of data and sensitivity to the WD mass. So it might be nice to find a shortcut.

Such a shortcut is provided by the ultraviolet photometric survey of GALEX. This provides snapshot FUV(1350–1750 A) and NUV(2150–2700 A) magnitudes for roughly half of all CVs. For our problem of estimating $T_{WD}$ in these faint CVs, this is quite useful, for two essentially fortuitous reasons:

(1) The temperatures certified by spectroscopy are mostly in the range 10000–16000 K, where the "quasi-molecular" satellite features of Lyman α (e.g., Allard & Koester 1992) provide a very sensitive thermometer. The cooler WDs in this range have virtually no flux below 1650 A, while the hotter stars lack these very broad absorptions. But the NUV band is smooth, and therefore tracks temperature very gently. Thus the FUV-NUV color is a sensitive probe of $T_{WD}$ in this range.

(2) Previous ultraviolet spectroscopy suggests that for intrinsically faint dwarf novae, the WD tends to dominate the UV spectrum. Therefore accretion-disk contamination is likely to be small, and the FUV-NUV color can be informative.[20]

We attempt to exploit this sensitivity in Figure 9. This shows FUV-NUV versus $P_{orb}$ for all measured dwarf novae, assuming: (a) the measurement error does not exceed 0.35 mag, and (b) the measured NUV magnitude indicates "near quiescence". The shape is consistent with the boomerang shape which betrays period bounce, and the "B" symbols indicate stars identified as bouncer candidates *on other grounds* (mainly residence in the nether regions of Figures 5 and 6). The temperature scale at right shows the values of $T_{WD}$ associated with this GALEX color in the models of Kawka & Vennes (2007), assuming a WD mass of 0.9 $M_o$. For the coolest WDs these values of $T_{WD}$ may be upper limits, because the FUV band in these stars is almost certainly contaminated by line emission (esp. C IV λ1550 and He II λ1640). No correction is applied in Figure 9; but the contaminating line emission is measurable in two stars (GD 552 and V455 And), and the arrow shows the size of the correction (Unda-Sanzana et al. 2008; Araujo-Betancor et al. 2005).

---

20 Although somewhat contaminated by unwanted continuum light from the disk. The quality of the clue depends on how clearly the WD dominates the UV (or optical) spectrum – as revealed by that spectrum, or by the star's $M_v$.

So this analysis appears to yield another boomerang, and appears to flag the same stars as the bouncer candidates.

## 13. WHITE DWARF TEMPERATURE VERSUS Q

In theory, $T_{WD}$ and $q$ should both decline monotonically with age, and hence should be highly correlated – in a manner somewhat resembling Figure 7. Figure 10 shows the analogue of Figure 7, with the FUV-NUV color standing in for $T_{WD}$. The FUV and NUV magnitudes are snapshots (with exposures often ~100 s) and are not strictly simultaneous. Interpreted as signatures of $T_{WD}$, they are additionally problematic – because they are contaminated by emission lines and disk continuum, and because $T_{WD}$ is very sensitive to $M_1$ (T~$M_1$, hence with a ±20% scatter arising merely from the inevitable scatter in $M_1$)[21]. Amid the noise, Figure 10 nevertheless shows that correlation. The apparent sharp drop around FUV-NUV = 0.6 corresponds to $T_{WD}$=12500 K, in the 0.8 $M_o$ models of Kawka & Vennes (2007). The likely range in $M_1$ (say 0.6-1.1 $M_o$) may well explain the large observed scatter in that region, because T is sensitive to $M_1$, and FUV-NUV should be very sensitive to T near 12500 K, where quasi-molecular Lyman-α absorptions start to annihilate the FUV flux in WD model atmospheres.

## 14. WHO, EXACTLY, ARE THE BOUNCERS?

### 14.1 The Idea

Period bouncers have been proposed as the final stage of CV evolution since 1981 (Paczynski & Sienkiewicz 1981; Rappaport, Joss, & Webbink 1982). Recent theoretical models (Kolb 1993, Kolb & Baraffe 1999, Baraffe & Kolb 2000; Howell, Nelson, & Rappaport 2001) suggest that a complete census of CVs should be dominated by these stars, since their evolution is very slow and prior evolution is fairly fast. Yet as recently as the 2004 Strasbourg meeting, Tom Marsh turned to the audience and asked: "Does anyone know the **names** of any of these stars?"... and was answered by silence.

Period bounce should occur when the secondary's thermal timescale becomes longer than the mass-transfer timescale. These times scale respectively as $M_2^{-1}$ and $M_2$, so they cross over (equalize) when $M_2$ is sufficiently low. When mass-transfer is driven by $\dot{J}$ from gravitational radiation (GR) alone, this occurs when $M_2$ is near 0.07 $M_O$ (Paczynski 1981, Rappaport et al. 1982). For single stars, this is a magic number, and that has sometimes been a source of confusion to observers (and referees). Many papers on this subject have announced the discovery of "degenerate secondaries". But a recent-period-bouncing secondary is significantly supported by ordinary thermal pressure; that's really the point – the secondary bloats due to excessive temperature (inability to cool fast enough)[22]. Observers have also been overly seduced by the issue of whether the secondaries are "substellar" (below the Kumar limit of ~0.075 $M_\odot$ for core H burning). This too misses the key point. The Kumar limit has no great significance in CV evolution; depending on the secondary's mass-radius relation, and any extra mechanisms which might enhance $\dot{J}$ over the GR rate, the cross-over of the two relevant timescales (more specifically, period bounce) occurs somewhere in the range 0.05–0.09 $M_\odot$[23] – and has no direct relation to the Kumar limit.

---

21 Luminosity scales as $R^2T^4$, and accretion luminosity scales as M/R; so if the energy source is accretion, then $T^4$ scales as $M/R^3$. The WDs in CVs are fairly massive (0.8-1.0 $M_o$), and in this domain R scales roughly as $M^{-1}$; so T scales roughly as M.

### 14.2 Properties and Candidates

So period bouncers should have low $M_2$, but the precise mass of a particular secondary does not signify its evolutionary status or history. Nor does a precise comparison to models, unless those models are independently known to be precisely correct. The best probe lies in empirical plots like Figures 4–10, because the correct theoretical curve should reproduce the observed general trends (rather than simply matching one star, or just a few, or matching the exact numerical predictions of a specific model[24]). There appears to be rough agreement: the observable quantities $<M_{ve}>$, $q$, and FUV-NUV all show a characteristic boomerang, as theory would predict. But a greater population on the lower branch would be very welcome, as would improvements in the theoretical understanding of mass-radius and the drivers of evolution (presumably angular-momentum loss).

The bouncers are then the stars which appear to be on the lower branch. This is best seen in Figures 6/8 and 9, and to some extent Figure 5. Residence at the lower left of Figure 7 also confers some credentials, though tarnished by the many limits in this region. An excellent candidate should have low $q$ (or $M_2$), cool $T_{WD}$, faint $<M_{ve}>$, and faint $M_{vq}$. A relatively long $P_{orb}$ also greatly builds the case for bouncer status, since the gap between lower and upper branches grows with increasing $P_{orb}$. Near minimum $P_{orb}$, this latter point somewhat muddies the case for nearly every star, including those previously labelled as period-bouncers from evidence described as "direct"[25]. We assemble a list of good candidates in Table 3, more or less in order of decreasing credential quality.[26]

---

22 Although the recently-bounced star moves to the degenerate branch fairly soon (see e.g. Figure 1 of Paczynski & Sienkiewicz 1981) if the driver is as weak as unassisted GR.

23 Or an even greater range, if more exotic assumptions are made about angular-momentum loss and the secondary's hydrogen content. Since $P_{orb}$ and $q$ at apparent period bounce (see Figure 6 and accompanying discussion) are *close* to the values mandated by GR and X=0.7, and since the observed curve is qualitatively a boomerang as predicted by theory, we conclude that observations rule out very exotic assumptions.

24 There is just too much uncertainty in the measurements, and too much idiosyncrasy in the stars, to rely on precision surgery here. Brute force is better.

25 A familiar word, but used in the astronomical literature with a technical meaning: Derived In Research Employing Collaborators' Techniques.

26 Actually, this list is prepared without consideration of $T_{WD}$. This is partly in order to give significance to the test we used in discussing Figure 8: the special location of the "B" stars. It is also partly because Of the suspects in Table 3, five are estimated to be within 100 pc, and seven within 120 pc.  Among all other dwarf novae, the corresponding numbers are two (VW Hyi and OY Car) and seven (VW Hyi, VY Aqr, OY Car, T Leo, U Gem, Z Cha, and IP Peg).  This simple nose-counting in the solar neighborhood suggests a roughly 60% representation among dwarf novae, and dwarf novae are probably the largest subset of CVs.  This is before any consideration of the selection effects which strongly discriminate against their discovery: their faintness and reluctance to erupt.  With so many good and nearby candidates, and the selection effects sure to suppress the count (and the percentage), the true percentage is clearly very high.

We discuss selection effects in the next section.

15. **RECURRENCE TIME, SCALE HEIGHT, SPACE DENSITY, AND COMPLETENESS**

Table 2 has many stars and distances – and with a few assumptions, could yield valuable information about the Galaxy's CV population.  But the assumptions are not innocuous.  These have been extensively discussed by Pretorius, Knigge, & Kolb (2007, hereafter PKK) and G09, with application mainly to the Palomar-Green survey and the SDSS.  The discussion of PKK, especially, underlines the sensitivity of survey results to limiting magnitude, color selection, and Galactic latitude.  The present study is somewhat different, though.  The majority (77%) of stars in Table 2 were discovered through traditional variable-star means – i.e., a human, or occasionally a computer, noticed an outburst.  This large component constitutes an additional "survey", the grandest of them all: the Survey of humans observing the night sky.  It's important enough to merit a name: the VHS, or Visual Historic (or Human) Survey.  Based on the human eye[27] and its interface with the brain, the VHS uses optical technology and computing power far beyond any of its competitors.  It has been running for >100 years, with little dependence on color or reddening or Galactic latitude.  Because the pattern of DN variability is so distinctive, it does not suffer from confusion with other common objects – quasars, white dwarfs, subdwarfs, other types of variable stars, etc.  These advantages simplify some potential complications; the main issues are brightness and recurrence time.

15.1  Recurrence Time

Of course, brightness depends on distance – and for the stars discovered as erupting dwarf novae, *only* on distance, since these stars always have $M_v \approx +5$.  The effect of recurrence time is more pernicious. In Figure 11 we use the data of Table 2 to explore the dependence of $T_{rec}$ on mass ratio.  Triangles indicate stars with a lower limit on $T_{rec}$ (with less than 2 eruptions observed, and with an observational record adequate

---

the FUV-NUV test has hazards: nonsimultaneous observation, recent and unrecognized eruptions, and contamination by accretion light.

27 And its venerable servant, the photographic plate.  Many thousands of variable stars were discovered by photography – but with a human eye (usually, two) "blinking" the plate.

to exclude lower values). Open circles indicate stars with a lower limit on $T_{rec}$ and upper limit on $q$. We fit the points (excluding the limits) with a straight line in log-log space, and find

$$T_{rec} = 318(\pm30) \text{ days } (q/0.15)^{-2.63\pm0.17} . \qquad (8)$$

So sharp a dependence on $q$ insures that humanity gets only rare opportunities to discover (through eruptions) the low-$q$ stars. Compared to a garden-variety DN with $q$=0.15, a period-bouncer candidate with $q$=0.05 erupts ~18x less often. Their eruptions last about twice as long; so, in round numbers, period-bouncers are ~10x less prominent on our radar screens of variability.

### 15.2 Galactic Scale Height

We used the distance estimate and Galactic latitude to calculate each star's height z above the Galactic plane, assuming the Sun to be exactly at the mid-plane. Figure 12 shows the distribution, along with an exponential distribution with a scale height $h$ = 300 pc. Twenty-six years ago, the same plot (Figure 9 of P84) showed an exponential scale height of 150 pc (similar to the quoted result of a Gaussian scale height of 190 pc). The population considered there was dominated by CVs of long $P_{orb}$, whereas the stars studied here are all of short $P_{orb}$. The difference in scale height probably arises because the short-$P_{orb}$ stars are systematically older (see PKK for a lucid discussion of this).

A formal fit is not significant, because stars are selectively missed at very low z (due to obscuration, confusion, and lack of search) and very high z (due to distance). A scale height of 300 pc fits the broad middle, and shows undercounts in both regimes where undercounts are expected. So 300 pc might be roughly correct, for the particular mixed-population of stars in our census.

### 15.3 Completeness and Space Density

The VHS is not well suited for the study of space density, because there is no definite (or even approximate) magnitude limit for detection. And by adding to it the 23% of stars found by other means, the "sample" is further polluted – quite seriously, since that 23% has rather different properties (many more low-luminosity systems, dominated by the SDSS CVs). After another decade of robotic surveys for variability, these issues might be greatly clarified. On the other hand, we have numerous distances, numerous $q$s, and empirical laws for $L_v(q)$ and $T_{rec}(q)$; these will permit us to make some progress on this issue.

To make some estimate of how many stars are missed, we ordered them into six categories by distance, and counted the number in each spherical shell, with the result given in Table 4. Assuming all-sky searches, no interstellar absorption, no stars missed, the Sun at the Galactic mid-plane, and a single CV population with a scale height of 300 pc, then the effective volume searches in each category imply space densities in the solar neighborhood given in the last column of Table 4. These are plotted in the upper frame of Figure 13. Since there is nothing special about the immediate solar neighborhood (Cusanus 1440, Bruno 1584), this curve should be flat. Its decline with distance proves that stars are missed due to distance (i.e., limiting magnitude); in particular, roughly 2/3 are seemingly missed at ~400 pc. This is easy to understand: at 400 pc, a dwarf nova erupts only to $V$=12.8, somewhat too faint to be efficiently found by serendipitous methods (visual observation, or the various photographic surveys of the past century). Given the sharp $T_{rec}(q)$ dependence of Figure 11, you would expect to miss predominantly low-$q$ stars, since they offer fewer chances for serendipitous discovery.

And you would be correct. The lower frame of Figure 13 shows the average value of $q$ in each spherical shell. It grows with increasing distance – again appearing to contradict Bruno, Nicholas of Cusa, etc. Low $q$, and consequently infrequent eruption, handicaps discovery at any distance; but the handicap is not as severe for nearby stars, since their eruptions may be bright enough to be caught by more efficient discovery methods (binoculars and wide-field astrographs). So this analysis suggests $N_0 \approx 1.8 \times 10^{-6}$ stars/pc$^3$.

This assumes that human vigilance is sufficient to catch them all in our immediate neighborhood, and to measure $P_{orb}$ (required for inclusion in the census[28]). In reality, the stars are too bashful, and the humans too ineffective, to satisfy this condition. The correct number must be significantly higher.

How much higher? Well, Table 2 contains 211 northern stars, and 81 southern. This imbalance arises mainly from the scarcity of land masses (and hence astronomers, cameras, and telescopes) in the Earth's southern hemisphere. So our estimate needs a geographic-bias correction of 1.5x – yielding $N_0 \approx 2.7 \times 10^{-6}$. Still missing are corrections for the stars' bashfulness (faintness and failure to erupt) and the humans' ineffectiveness (failure to notice, and to measure $P_{orb}$). So, in the fullness of time and energy, another factor of a few seems plausible – and even likely, given the success of SDSS in finding faint noneruptive CVs.

Can the number approach $10^{-4}$ pc$^{-3}$, as predicted by some theories? Probably not. This question has been carefully addressed on three occasions. G09 discussed the SDSS; PKK discussed several, and especially the Palomar-Green survey; and P84 discussed a half-dozen surveys, as well as the census of CVs presenting themselves to us in miscellaneous ways (usually through eruption). The conclusions were similar. In the language of P84 (see § V and Figures 10/11 of that paper), *that space density floods the sky – and especially the immediate solar neighborhood – with vast numbers of CVs which should be observed, and are not.* Nearby CVs should be found in surveys based on blue color, emission lines, eruptions, and X-rays; and the category we have alleged as the most populous – the "WZ Sge stars" – prominently displays *all* of these properties. To hide from the surveys, a large population of nearby CVs must avoid all these radar screens of detection, and hide in the thicket of a much larger population (probably solitary WDs, with $N_o = 10^{-2}$ pc$^{-3}$). This excuse is still viable[29], but requires mighty careful engineering.

---

28 On the grounds that little is known about a binary until you know $P_{orb}$. For the limited purpose of learning the space density, this is not quite so true; a study of the distances of "unknown $P_{orb}$" CVs could be interesting.

## 16. PERIOD-BOUNCERS: THE EXPANDED ROSTER

Except at relatively long $P_{orb}$, observations do not cleanly separate stars into pre-bounce and post-bounce – and this is unexpected anyway, since there would be a "missing link" difficulty. But we can assign a score based on a star's evidence of characteristics expected from bouncers on theoretical and empirical grounds. In roughly descending order of importance, these are: low $q$ (or $M_2$), cool $T_{WD}$, faint $<M_{ve}>$, long $P_{orb}$ relative to the 76 minute minimum (only a credit if a star has other bouncer hallmarks), faint $M_{vq}$, long $T_{rec}$, large DN amplitude, long DN plateau. Some of these properties are obviously correlated (e.g., $q$ versus $<M_{ve}>$ and $T_{rec}$, DN amplitude versus $M_{vq}$), but it's important to have a long list of tests, because the observational record is fragmentary, and we ought to use all available clues.

Credit is assigned on two other grounds. Several stars have very high proper motion, suggestive of great age, and several others show signs of a substantial 2:1 resonance in the accretion disk – likely available only to stars of low $q$. These are not necessarily qualities we would "expect" of a period-bouncer. But if present, they are points with some evidentiary value. So we credit them under *other*.

Table 5 presents a liberal list of candidate bouncers, again in roughly descending order of overall credential quality. Each Y means a credit, each – means "no information", and each I means indifferent (quantity measured, but not really distinctive, or not sufficiently precise). Y? is a slight variation on I: measured and distinctive, but not with very high confidence (e.g. a weak detection). Each YY or YYY is a double or triple credit, either because the measured property is critical ($q$, $T_{WD}$, $<M_{ve}>$) or because the measured number is very distinctive, or both. Each N means "measured and *un*characteristic of period-bouncers", and is counted as minus one Y. Score the number of Ys, and look for the big numbers. Those are the stars you want on your team.[30]

---

29 Considering how little we understand about DN quiescence, it's admittedly a significant excuse – at quiescence. In fact, quite a few stars of *nearly* this description were found by the SDSS (G09). But if these *incognito* stars ever erupt, their nondetection is puzzling. Assuming $T_{rec}$ = 50 yr, 50% sky coverage, and 50% time coverage, then $N_o = 10^{-4}$ implies that 8 years of ASAS-3 should have detected ~20 of these stars very clearly ($V<10$). The observed number seems to be 1 (GW Lib in 2006). It is also interesting that the only one found (at that brightness limit) was a previously known star; this suggests that there is no vast armada of similar nearby and eruptive stars waiting to be discovered.

30 Although it is only #5 on our list, WZ Sge – with its very long $T_{rec}$ and puny secondary – should by historical precedent be considered the team captain. As the nearest of all known CVs, it's also a good choice for captaincy of a very populous class. But it may lose that status (likely to GD 552) if greater

Data are fragmentary on many of these stars, but we reckon that the top 20 on the list are good candidates (by our subjective overall judgment, which is not exactly the same as the numerical score). All are decent candidates, however.

How do normal short-period dwarf novae score? Well-studied stars in this category are usually a clean sweep of Ns, and hence around -10. We randomly selected three, and add them to the bottom of Table 5 for comparison. Quite possibly there is just *one* physical quantity which determines all of these other observational properties. If so, a good guess is *q*, which does a good job of distinguishing the upper and lower branches (Figure 6). (This probably really means $M_2$, but that quantity is usually far less accessible to observation.)

## 17. SUMMARY AND THE VIEW AHEAD

(1) We collect available data on 46 stars which are suitable for calibrating the "standard candle" of a dwarf-nova outburst. We derive a new relationship between $(M_v)_{max}$ and $P_{orb}$ – very similar to Warner's 1987 relation – and suggest that the *plateaus* of DN light curves are also a useful, and sometimes more observable, clue to a star's distance. Researchers in the CV community have been oddly bashful about utilizing two important standard candles: that of the DN outburst, known since 1965 – and that of the WD, essentially known since Alvan Clark first gazed at Sirius B on 31 January 1862 [or more specifically since 1957, when Greenstein first identified the WD in WZ Sge (Greenstein 1957)]. Get used to it: *distances are not that difficult to estimate*.

(2) We have compiled a large catalogue of known or likely dwarf novae with $P_{orb}$ < 3 hours. Adding the DN standard-candle relation [more precisely, Eqs. (4)-(6)] to the armada of other constraints, we estimate distance, $M_v$ at quiescence, and time-averaged eruptive $M_v$. The results demonstrate that accretion luminosity in short-period CVs depends on $P_{orb}$, with a scatter of ~1.0-1.5 mag. The distributions of $M_{vq}$ and <$M_{ve}$> versus $P_{orb}$ (Figures 4 and 5) show that CVs become much fainter as they approach minimum $P_{orb}$. Both <$M_{ve}$>($P_{orb}$) and $q(P_{orb})$ show a "boomerang" shape. This is an expected signature of period bounce, if the stars flagged as highly evolved (period bouncers) in the two distributions are the same stars. Figure 7, plotting <$M_{ve}$> versus $q$, establishes that they *are* the same stars: CVs of very low $q$ are very faint, and vice versa. These are the period-bouncer suspects. Such bashful stars are very rare erupters and very faint in quiescence, and hence quite difficult to discover. Nevertheless, a census of the leading suspects makes it clear that this is a very common type of CV, probably comprising >50% of the entire DN population.

---

emphasis is placed on its relatively high $T_{WD}$. Although $T_{WD}$ furnishes an important clue, there are uncertainties surrounding its interpretation: WD mass, time since the last DN eruption, time since the last CN eruption, and even binary inclination (courtesy of the disk geometry, which could render the WD equatorial regions somewhat hotter). This, added to all the other reasons, is why *all* observational clues should be utilized.

(3) Ultraviolet colors also do a good job of flagging bouncers, due to their great sensitivity to temperature in the range of interest (10000–13000 K). These also trace out a boomerang curve, and the same stars tend to be on the lower branch.

(4) Along the way, we study the proper motions of 141 CVs (not just dwarf novae) with adequate distance estimates. As a class they suggest a characteristic $v_{tang}$ near 40 km/s. The bouncer suspects appear to be moving faster, with $v_{tang}$ near 55 km/s. If true, this could mean that period bouncers represent an older population.

(5) We use Table 2 to study the dependence of DN recurrence time on mass ratio, and find $T_{rec} \sim q^{-2.6}$. This underlines and quantifies the difficulty in finding stars of low $q$: they seldom erupt (with $T_{rec}$ = 20 years in the period-bouncer domain near $q$=0.05).

(6) We use our distances to study the distribution of short-period dwarf novae in the solar neighborhood. Reckoned as a single population, the stars have an exponential scale height (above the Galactic plane) of 300±60 pc. This compares to 150 pc in an earlier study, dominated by stars of long $P_{orb}$. As discussed by PKK, the true population is surely a mix, with the scale heights dependent on age – just like other stars in the Galaxy.

(7) Space density is more troublesome. By counting the number in spherical shells of increasing volume, we can estimate the undercount relative to the immediate solar neighborhood (~120 pc). Out to 400 pc, we apparently undercount by a factor ~3; Figure 13 shows that this is due to selective undercount of low-$q$ systems (mainly because they seldom erupt). The data suggest a short-period DN space density ~2.7 x $10^{-6}$ stars/pc$^3$. But this counts only stars which have appeared on our various radar screens (mainly through eruptions) and have yielded $P_{orb}$. Most of these low-$q$, seldom-erupting, faint-$M_v$ binaries probably still elude discovery.

(8) It is hard – practically impossible – to imagine how distributions like Figures 4-10 could be made without an evolutionary effect like period bounce. But the most interesting domain is at low $q$, where the accuracy (of Figure 6/8, the most significant curve) is lower. The $q$ limits from radial-velocity studies require adopting limits for $K_1$ and $i$, which might not be sufficiently conservative. Some of the $P_{orb}$ measures come from transient photometric waves, with a duration too short to yield small error bars (required for a good-quality ε). And most of the points come from superhump studies; these are all subject to possible inaccuracy of the ε($q$) calibration at low $q$, and a few have $P_{sh}$ values not known with certainty to comply with our standard (4 days after superhump onset). These issues of accuracy could significantly, though not qualitatively, change the appearance of the lower branch of Figure 6. More seriously, a few values of $P_{orb}$ come from a photometric wave which is *weak* (say 0.02 mag) as well as short-lived. Those detections are only probable, not certain (and are flagged in Table 2 with a ?). Clarification of all these points in future work would greatly help in defining the lower branch. However, it's reassuring that Figures 5 and 9, which are independent of *q* and of each other, seem to convey the same message as Figure 6.

(9) In arriving at these results about nonmagnetic CVs with $P_{orb}$ < 3 hrs, we had to suppress some complications. First, we excluded stars which suffered a recent classical-nova eruption. That seems fair: the stars are likely still too contaminated by their recent thermonuclear runaway. Second, we excluded stars which otherwise resemble classical-nova remnants (high-excitation spectra, bright $M_v$, no dwarf-nova eruptions), but which are not associated with any historical nova. This exclusion is only slightly more embarrassing, since there are very few such stars, and since we have only been recording (or more accurately, recording and remembering) classical novae for ~100 years.[31] Both of these exclusions can be generically justified by the limitation of this study to "known and likely dwarf novae". The third exclusion is more embarrassing: the ER UMa stars. These are certainly *bona fide* dwarf novae, but their luminosities are far above the trend line in Figure 7 – typically with $<M_{ve}> \approx$ +7.2. We take some solace in the rarity of these stars (5 of the 270 stars tabulated, and probably <1% of the population in space density). But the main reason for their (partial) suppression is simply that we don't yet understand how stars of low *q* can be so luminous. This remains a major flaw in our understanding of CV evolution.

(10) It would be fascinating to know whether these three exclusions really amounted to just one, based on time since the last classical-nova eruption. Current understanding of dwarf novae has an odd feature: we're pretty sure, based on the known physics of hydrogen in a degenerate environment, that all these H-rich dwarf novae are basically in the long interval between CN eruptions... yet no actual CN has been observed to resume life as a normal dwarf nova (although they often stabilize as something like a "UX UMa star" in 10-40 years). Maybe the stars recover from their nova jolt by evolving along a path like this: CN → supersoft binary in 1 year → UX UMa in 10 years → ER UMa in 1000 years → SU UMa in 10000 years → back to CN in $10^6$ years. That would be a simple and appealing story, not even requiring a change in constellation. By studying the space densities of these classes, we could investigate what timescales are needed for such a progression, and whether they make demographic and physical sense. It would be a lot nicer to have just one incompleteness in our story of short-period CV evolution, rather than three.

(11) But there are other flaws, too. A star can cool no faster than its thermal timescale, and this implies that period-bounce is *inevitable*, since GR exists as a minimal driver of evolution. The popular theory of late CV evolution [unassisted GR, first formally proposed by Paczynski (1981)] brought that realization to its deserved prominence; now it needs to score some successes on more specific tests. Figure 6 shows that it doesn't do so well in predicting the minimum $P_{orb}$, nor the value of *q* at a given $P_{orb}$. Figure 7 doesn't look promising either: it shows $L_v$ rising smoothly with *q*, whereas pure-GR theories predict an accretion rate almost flat with *q* on the upper branch of evolution

---

31 And one of these stars has even been associated with a possible ancient nova (BK Lyn with the guest star of 101 A.D.; Hertzog 1986).

(e.g. Figure 1 of Barker & Kolb 2003, Kolb & Baraffe 1999). These important problems still await solution.

(12) Humans like novelty, and in recent years the SDSS has been the "new kid on the block". Many, many authors have led cheers for the application of the SDSS to these issues of CV population and evolution. Follow-up observation of SDSS CVs with large telescopes has led to substantial progress, reported in many papers and recently summarized (G09, L08). In contrast, stars from the VHS (Visual Historic Survey) have been relatively neglected – often relegated to study with amateur telescopes. Here we praise, and describe explicitly in §15, the advantages of the VHS. It forms the main basis for this study, although we include many SDSS stars and a smattering of stars discovered in other ways (e.g. PG and Rosat surveys). This omnivorous diet endows our study with *bulk*, and takes advantage of eruptions, the highly important property that puts the "C" in CVs. This eases the task of seeing trends in these figures, despite the many contributions to scatter (errors in distance, *V*, inclination, and $T_{rec}$, and the likely dispersion in WD mass). It also minimizes the chance of significant contamination by stars which are *incorrectly classified* as CVs. Both approaches (SDSS and ours, which could be called "VHS+") are likely to improve a lot, when newly discovered orbital periods add significantly to the rosters.

(13) Surveys for optical transients (e.g. CSS, LSST, Palomar Transient Factory) will surely find many of the period bouncers, but saturation effects will likely blind them to the nearby erupters, which could be the most important for understanding the class. Wide-field surveys with *small* telescopes might address this problem better.

(14) This paper's neglect of *magnetic* CVs deserves another emphasis. Probably ~10-30% of all CVs are strongly magnetic, and very little in this paper applies to them. Norton et al. (2008) give a lucid account of their possible evolution.

(15) Distance may be the *sine qua non* of astrophysics, but accretion rate is the *ne plus ultra* of binary evolution. And we have not been able to extend our study quite that far. There are still major tasks ahead: convert $M_{vq}$ and $<M_{ve}>$ to bolometric luminosity, and learn how to convert from $L_{bol}$ to the actual accretion rate.

Many thanks to Jonathan Kemp, Tyler Harris, Jim Applegate, Stephan Vennes, Dinara Leshunou, and Chris Peters for conversations, technical assistance, and permission to cite results in advance of publication. And to John Thorstensen, who can always be relied on to play the "good cop" on issues of distance-finding. This research has been supported by generous grants from the NSF (AST-0908363), NASA (GO11621.03A), and especially the Mount Cuba Observatory Foundation (07-1605).

FIGURE CAPTIONS

Figure 1. Corrected $(M_v)_{max}$ versus $P_{orb}$ for the stars of Table 1. The top frame assumes a uniform dimming of +0.8 mag in translating the maximum of superoutbursts to that of normal outbursts. The linear fit is Eq. (3). The bottom frame applies no such correction, and the linear fit is Eq. (4). All data adopts the standard of "a typical bright eruption".

Figure 2. Proper motion versus distance$^{-1}$, for 141 CVs, of which 20 are candidate period-bouncers (B). The line shows a fit to the 121 other CVs, and corresponds to $v_{tang}$ = 39 km/s. A typical error on µ is 7 mas/yr. The bouncers appear to prefer the upper part of the figure, consistent with faster motions – unless their distances are systematically overestimated.

Figure 3. The equivalent width of H$\beta$ emission versus $M_v$, for nonmagnetic CVs with primary or secondary distance clues. Both quantities have been corrected for the presence of nonaccretion light (secondary or WD), but no account is made for the effect of binary inclination. Filled circles show stars with $P_{orb}$ < 3 hr, and open circles are stars with $P_{orb}$ > 3 hr. In addition to the stars shown, we could show – but don't – the hundreds of erupting dwarf novae with measurable values, all in the neighborhood of $M_v$ = 3–6, EW = 0–15 A.

Figure 4. Quiescent $M_v$ (accretion light only) versus $P_{orb}$. Stars within 100 pc are given a larger symbol, to suggest what a complete count might more closely resemble (viz., containing many more intrinsically faint stars). This convention applies also to Figures 5-8. There are 7 such stars; but for Figure 4 alone, deep eclipses disqualify one (OY Car). Points have an estimated observational error of ±0.9 mag.

Figure 5. Corrected average eruptive $M_v$ (<$M_{ve}$>, or "accretion $M_v$") versus $P_{orb}$. In a few cases, this is obtained by integrating under the eruption light curve. More commonly, it is calculated by the "square-wave" approximation described in § 9. Triangles are upper limits on <$M_{ve}$>, obtained from methods discussed in § 9 (pertinent to stars with an unknown recurrence time). Points have an estimated observational error of ±0.8 mag.

Figure 6. Mass ratio $q$ versus $P_{orb}$. Dots are positive measurements from eclipsers and superhumps; triangles are upper limits on $q$ from radial-velocity studies. The curve is the predicted trend if CV evolution is driven by angular-momentum loss at the gravitational-radiation (GR) rate, assuming X=0.7 and $M_1$=0.8 $M_o$. The error in $q$ is likely to be ~15% on the upper branch (say $q$>0.10), and somewhat higher on the lower branch.

Figure 7. <$M_{ve}$> versus $q$. Dots are measurements, and triangles show upper limits (to the brightness of <$M_{ve}$>, arising from the observed limits on DN recurrence time. Double arrows show upper limits on both $q$ and <$M_{ve}$>. The line corresponds to $L_v$ ~ $q^{2.4\pm0.2}$.

Figure 8. An amended version of Figure 6, where the open symbols indicate 41 added stars for which the value of $q$ is not measured, but inferred from the value of $<M_{ve}>$ (assuming they lie on the trend line of Figure 7). Open triangles come from upper limits on the accretion light (usually arising from the observed limit on the DN recurrence time).

Figure 9. FUV-NUV color versus $P_{orb}$. GALEX data is accepted if the magnitudes suggest quiescence, and if the FUV-NUV error does not exceed 0.35 mag. A few stars are added which lack GALEX data, but have FUV and NUV fluxes well-determined from HST spectroscopy. The "B" symbols show stars suspected to be period bouncers on other grounds (mainly $q$ and $<M_{ve}>$, and listed in Table 3). The FUV band of the reddest (coolest) stars are likely to be significantly contaminated by emission lines; no correction is applied, but the arrow shows the amount of the needed correction for one star (GD 552, Patterson et al. 2010). The scale at right shows the WD temperatures corresponding to these GALEX colors – assuming a WD mass of 0.9 $M_o$ (Kawka & Vennes 2007), and no contamination by accretion-disk continuum light.

Figure 10. FUV-NUV color versus $q$. Only dynamical $q$s are used (no surrogates from $<M_{ve}>$), and a few of the colors are synthesized from HST or IUE spectra. Interpreted as a signature of $T_{WD}$, the apparent change in slope around ($q$=0.07, FUV-NUV = 0.6) corresponds to $T_{WD}$=12500 K, right where quasi-molecular Lyman-α absorption starts to strongly affect the theoretical colors (see the scale in Figure 9).

Figure 11. Recurrence time versus mass ratio. Filled circles are measurements, and open circles are lower limits to $T_{rec}$ (when fewer than 2 outbursts are observed despite a lengthy observing record). Open triangles are lower limits to $T_{rec}$ and upper limits on $q$, and the line is Eq. (8). The inevitable 3-5 month gaps between observing seasons inflict uncertainty on all estimates of $T_{rec}$, and this is particularly severe for the rarest erupters.

Figure 12. The number of stars as a function of height above the Galactic plane, compared to an exponential distribution [exp(-$z/h$)] with $h$ = 300 pc.

Figure 13. *Upper frame*, the space density of short-period dwarf novae in the volumes defined in Table 4, under the assumptions described in the text. The decline with increasing distance arises from a severe undercount. *Lower frame*, the average $q$ in these volumes. This shows specifically that the undercounted (and undiscovered) stars at large distance are predominantly of low $q$. Reminder: these statistics are limited to known or likely dwarf novae with known or estimated $P_{orb}$<3 hr. Magnetism, long $P_{orb}$, unknown $P_{orb}$, and lack of discovery are probably all important categories – and could easily boost the mid-plane space density of all (active) CVs to near $10^{-5}$ pc$^{-3}$.

TABLE 1 – DWARF NOVA CALIBRATORS

| Star | $P_{orb}$ (d) | $V_{max}$ | $d$ (pc) | $M_v$ | $i$ | $\Delta M_v$ | $(M_v)_{corr}$ | References |
|---|---|---|---|---|---|---|---|---|
| BV Cen | 0.6101 | 10.9→10.8 | 400 | 2.8 | 53 | 0.15 | 2.9(6) | Gilliland 1982, Watson et al. 2007 |
| UY Pup | 0.4793 | 13.0→12.2 | 1300 | 1.7 | 15 | 0.93 | 2.6(8) | Lockley et al. 1999, Thorstensen et al. 2004 |
| EY Cyg | 0.4593 | 11.9→11.6 | 1000 | 1.6 | 14 | 0.94 | 2.5(7) | Echevarria et al. 2007a |
| DX And | 0.4405 | 11.3→11.1 | 630 | 2.1 | 45 | 0.41 | 2.5(5) | Drew et al. 1993, Bruch et al. 1997 |
| AT Ara | 0.3755 | 11.5→11.2 | 630 | 2.2 | 38 | 0.60 | 2.8(8) | Bruch 2003 |
| RU Peg | 0.3746 | 10.0→9.9 | 282* | 2.6 | 40 | 0.55 | 3.1(5) | H04, Stover 1981 |
| CH UMa | 0.3432 | 11.0→10.8 | 480 | 2.4 | 21 | 0.88 | 3.3(6) | Thorstensen et al. 2004, Simon 2000 |
| MU Cen | 0.342 | 11.8→11.6 | 370 | 3.8 | 60 | −0.14 | 3.6(7) | Friend et al. 1990 |
| AF Cam | 0.324 | 13.3→12.3 | 850 | 2.7 | 45 | 0.41 | 3.1(8) | Thorstensen & Taylor 2001 |
| EM Cyg | 0.2909 | 12.3→12.0 | 400 | 4.0 | 67 | −0.55 | 3.4(5) | North et al. 2000 |
| Z Cam | 0.2898 | 10.2→10.2 | 163* | 4.1 | 62 | −0.25 | 3.8(6) | T03, Kraft et al. 1969 |
| SS Cyg | 0.2751 | 8.3→8.3 | 165* | 2.2 | 50 | 0.30 | 2.5(6) | H04, Bitner et al. 2007 |
| TT Crt | 0.2684 | 12.7→12.6 | 600 | 3.7 | 60 | −0.14 | 3.5(6) | Thorstensen et al. 2004 |
| BV Pup | 0.265 | 13.1→12.9 | 600 | 4.0 | 65 | −0.40 | 3.6(8) | Szkody et al. 1986 |
| AH Her | 0.2581 | 11.5→11.4 | 450 | 3.1 | 40 | 0.55 | 3.6(7) | Spogli et al. 2006, Horne et al. 1986 |
| SS Aur | 0.1828 | 10.6→10.4 | 201 | 3.9 | 38 | 0.60 | 4.5(6) | Sion et al. 2008, H04, Shafter & Harkness 1986 |
| RX And | 0.2099 | 10.6→10.6 | 200 | 4.1 | 55 | 0.07 | 4.0(6) | Shafter 1983, K06, Sion et al. 2001 |
| BD Pav | 0.1793 | 12.5→12.5 | 420 | 4.4 | 68 | −0.60 | 3.8(7) | Barwig & Schoembs 1983, Sion et al. 2008 |
| U Gem | 0.1769 | 9.1→9.1 | 102* | 4.1 | 70 | −0.72 | 3.4(4) | H04, Echevarria et al. 2007b |
| CW Mon | 0.1766 | 12.3→12.1 | 280 | 4.9 | 67 | −0.55 | 4.3(7) | Szkody & Mateo 1986 |
| CN Ori | 0.1632 | 11.8→11.7 | 400 | 3.7 | 45 | 0.41 | 4.1(7) | Friend et al. 1990, K06 |
| AR And | 0.1630 | 12.1→12.0 | 300 | 4.6 | 40 | 0.55 | 5.1(7) | Taylor & Thorstensen 1996 |
| KT Per | 0.1627 | 11.8→11.7 | 180 | 5.4 | 60 | −0.14 | 5.2(7) | T08, Thorstensen & Ringwald 1997 |
| IP Peg | 0.1582 | 12.2→12.2 | 115 | 6.9 | 84 | −2.3 | 4.6(7) | Smak 2002, Beekman et al. 2000 |
| J1702+32 | 0.1001 | 13.6→14.4 | 320 | 6.9 | 80 | −1.7 | 5.2(9) | Littlefair et al. 2006 |
| IR Com | 0.0870 | 13.5→14.3 | 300 | 7.0 | 80 | −1.7 | 5.3(7) | Feline et al. 2005, Kato al. 2002 |
| YZ Cnc | 0.0868 | 11.0→11.7 | 260* | 4.6 | 40 | 0.55 | 5.1(6) | Shafter & Hessman 1988, H04 |
| DV UMa | 0.0858 | 14.7→15.8 | 350 | 8.1 | 84 | −2.3 | 5.8(9) | Feline et al. 2004a, Patterson et al. 2000 |
| SU UMa | 0.0763 | 11.2→12.1 | 260 | 5.1 | 42 | 0.50 | 5.6(6) | Thorstensen et al. 1986, T03 |
| V893 Sco | 0.0759 | 12.7→12.6 | 155* | 6.6 | 75 | −0.9 | 5.7(6) | T03, Mason et al. 2001 |
| Z Cha | 0.0745 | 11.9→12.7 | 110 | 7.5 | 82 | −1.9 | 5.6(7) | Beuermann 2006, Wade & Horne 1988 |
| VW Hyi | 0.0743 | 8.5→9.3 | 75 | 4.9 | 57 | 0.00 | 4.9(5) | Sion et al. 1995, Schoembs & Vogt 1981 |
| IY UMa | 0.0739 | 13.0→13.9 | 190 | 7.5 | 86 | −2.6 | 4.9(7) | Steeghs et al. 2003, Rolfe et al. 2005 |
| HT Cas | 0.0736 | 12.2→13.1 | 123* | 7.7 | 81 | −1.8 | 5.9(6) | Feline et al. 2005, T08 |
| OU Vir | 0.0727 | 14.5→15.3 | 450 | 7.0 | 79 | −1.5 | 5.5(7) | Feline et al. 2004b |
| BZ UMa | 0.0680 | 10.6→11.3 | 220 | 4.6 | 57 | 0.00 | 4.6(6) | T08, Gansicke et al. 2003 |
| OY Car | 0.0631 | 12.1→13.0 | 100 | 8.0 | 83 | −2.0 | 6.0(7) | L08, Wood et al. 1989 |
| VY Aqr | 0.0631 | 10.1→10.9 | 97* | 5.9 | 60 | −0.14 | 5.7(6) | T03, Thorstensen & Taylor 1997 |
| J1227+51 | 0.0630 | 14.7→15.5 | 380 | 7.6 | 84 | −2.2 | 5.4(9) | L08 |
| V2051 Oph | 0.0624 | 12.2→13.0 | 160 | 7.0 | 83 | −2.0 | 5.0(8) | Baptista et al. 1998 |
| XZ Eri | 0.0612 | 14.6→15.4 | 400 | 7.4 | 80 | −1.7 | 5.7(8) | Feline et al. 2004a |
| T Leo | 0.0588 | 10.0→10.8 | 101* | 5.8 | 55 | 0.07 | 5.8(6) | T03, Shafter & Szkody 1984, Hamilton & Sion 2004 |
| SW UMa | 0.0568 | 10.2→11.0 | 165* | 4.9 | 45 | 0.41 | 5.3(5) | T08, Gansicke et al. 2005 |
| WZ Sge | 0.0567 | 8.3→9.1 | 43* | 5.9 | 77 | −1.2 | 4.7(4) | H04, Patterson et al. 2002 |
| V455 And | 0.0563 | 9.0→9.8 | 74* | 4.7 | 70 | −0.7 | 4.0(4) | Araujo-Betancor et al. 2005, G09 |
| GW Lib | 0.0533 | 8.8→9.7 | 100* | 4.7 | 13 | 0.9 | 5.6(6) | T03, Copperwheat et al. 2009 |

NOTES:
(1) In addition to the specialized references, we also compared, for all stars, the observed $K$ magnitude to the $M_K$ predicted by the Beuermann (2006) expression for surface brightness and the K06 mass-radius relation. This is essentially a "modified Bailey relation". For stars with prominent secondaries in the spectrum, or red turn-ups in the flux distribution, we considered the result to be a valid distance measure. For other stars, it gives only a limit.
(2) Asterisks indicate relatively precise (<15%) distances from astrometry. Other distances are rounded to the nearest 10 pc, to remind of their dependence on astrophysical assumptions, and/or quality (mostly 15-25%). Even in the best cases, the resultant total error in $(M_v)_{corr}$ is ~0.4 mag.
(3) Many stars below the period gap have few or poorly observed normal maxima. Therefore, for all stars with $P_{orb}$<0.12 d, we always measure the supermaxima, and then add 0.8 mag to obtain $V_{max}$

TABLE 2 – KNOWN AND LIKELY DWARF NOVAE WITH $P_{orb} < 3$ HR

| (1) | (2) | (3) | (4) | (5) | (6) | (7) | (8) | (9) | (10) | (11) |
|---|---|---|---|---|---|---|---|---|---|---|
| Star | $V_{max} \to V_{min}$ | $P_{orb}$ (d) | Type | $q$ | $\mu_x$ | $d$ (pc) | Square Wave | FUV | $(M_v)_{qui}$ | References |
| Coords | $(V_{max})_{qui}$ | $P_{sh}$ (d) | $P_{rec}$ (d) | | $\mu_y$ | $A_v$ | ($V$, days) | NUV | $<M_{ve}>$ | |
| | | $\ell + b$ | $i$ (°) | EW (Hβ) | $\gamma$ | Clues | | $T_{wd}$ (K) | $<M_{ve}>_{corr}$ | |
| BE Oct | 15.4→20.0 | | SU | | | 1100±250 | | 20.5 | (9.4) | Kemp & Patterson 1996, |
| 0000-77 | | 0.0771 | | | | 0.3 | | 19.6 | | Kato et al. 2003, |
| | | 307−39 | 30 | 86 | | 4,6 | | | | CBA, Mason & Howell 2003 |
| V402 And | 15.3→20.3 | | SU | | | 1100±250 | 15.9 | | 9.9 | Antipin 1998, K09, CBA |
| 0011+30 | | 0.0634 | 400 | | | 0.2 | 14 | | 8.8 | |
| Antipin V62 | | 113−32 | | | | 4 | | | 9.0 | |
| V592 Cas | 12.5 | 0.11506v | UX | 0.248 | −10 | 350±90 | | – | | Taylor et al. 1998 |
| 0020+55 | | 0.1222 | | | −14 | 0.4 | | | 4.5 | |
| | | 119−6 | 40 | 3 | −31 | 4,6,5 | | | 4.9 | |
| ASAS | 10.5→17.5 | 0.05604v | WZ | 0.097 | −74 | 160±40 | 11.3 | 18.50 | 11.6 | Templeton et al. 2006, CBA, |
| 0025+12 | | 0.0572 | >2000 | | −35 | 0.1 | 19 | 17.82 | >10.3 | Ishioka et al. 2007, K09, |
| | | 113−50 | 60 | | | 1,4,5,6 | | | >10.3 | Byckling et al. 2010 |
| EN Cet | 14.5→20.6 | 0.059p? | DN | | 8 | 800±250 | | >22 | (11.1) | Esamdin et al. 1997, |
| 0027−01 | | | | | −6 | 0.1 | | 21.5 | | Dillon et al. 2008, |
| | | 109−63 | 60 | 70 | | 4,6 | | | | Szkody et al. 2005 |
| KP Cas | 14.0→20 | | SU | | 12 | (600)±180 | 14.3 | – | (10.3) | Boyd et al. 2010, K09, CBA, US06 |
| 0037+61 | | 0.0853 | | | 0 | 0.7 | >8 | | | |
| | | 121−1 | | | | 4 | | | | |
| LL And | 13.6→19.8 | 0.05505p | SU | 0.131 | 6 | 550±150 | 14.2 | 20.6 | 11 | P03, K09, Howell et al. 2002 |
| 0041+26 | | 0.0567 | 1500? | | −10 | 0.2 | 18 | 20.5 | 9.8 | |
| | | 120−36 | 45 | 40 | | 3,4,5,6 | | 14300 | 10.1 | |
| OT | 14.7→22.8 | 0.0560o? | SU | 0.114? | | 1000 | 15.5 | 24.1 | 12.6 | Kasliwal et al. 2008, K09 |
| 0042+42 | | 0.0569 | | | | 0.2 | 21 | 23.3 | | |
| | | | | | | 4 | | | | |
| SDSS | 19.9 | 0.0572v | NL | | 27 | 500±150 | | 22.3 | 12.0 | Southworth et al. 2008, |
| 0043−00 | 20.4 | | >1000 | | 11 | 0.1 | | 20.7 | | Szkody et al. 2004 |
| | | 119−63 | 20 | 40 | | 3 | | | | |
| GX Cas | 13.5→19 | 0.0890v | SU | 0.19 | 6 | 450±140 | 13.9 | – | 10.4 | P03, K09 |
| 0049+56 | | 0.0940 | 160 | | −17 | 0.35 | 12 | | 7.8 | |
| | | 123−5 | 55 | 38 | −31 | 4,6,5 | | | 7.8 | |
| SDSS | 20.4 | 0.0557p | NL | | −10 | | | 21.55 | | Southworth et al. 2007, |
| 0050+00 | | | >600 | | −13 | 0.1 | | 21.1 | | Szkody et al. 2005 |
| | | 123−63 | 50 | 25 | | | | | | |
| V452 Cas | 15.2→19.5 | 0.0846p | SU | 0.206 | | 900±240 | 15.9 | – | 8.9 | CBA, Liu & Hu 2000, |
| 0052+54 | | 0.0888 | 146 | | | 0.8 | 12 | | 7.7 | Shears et al. 2009, P05 |
| | | 123−9 | | | | 4 | | | 7.9 | |
| HT Cas | 12.4→17.3E | 0.07365e | SU | 0.147 | 30 | 123±15 | 13.2 | – | 12.4 | Thorstensen et al. 2008, |
| 0110+60 | 17.9 | 0.0761 | 5000? | | −12 | 0.1 | 4 | | 15.1E | Feline et al. 2005, |
| | | 125−2 | 81 | 130 | −5 | 1,2,3 | | 13200 | 13.0E | Wood et al. 1992, Zhang et al. 1986 |

NOTE: The complete table of 292 stars is available at http://cbastro.org/dwarfnovashort/, along with a full explanation of each column (which is also given in § 7 of this paper).

TABLE 3 – PERIOD-BOUNCER CANDIDATES (NONMAGNETIC)

| Star | $P_{orb}$ (d) | $d$ (pc) | $q^1$ | $M_{vq}{}^2$ | $<M_{ve}>^3$ | $<M_{ve}>_{corr}{}^4$ | $T_{WD}{}^*$ (× 1000 K)$^5$ | References |
|---|---|---|---|---|---|---|---|---|
| GD 552 | 0.0713 | 74 | <0.052$v$ | 13.2 | | >11.7 | 10.5$gsd$ | Unda-Sanzana et al. 2008, Patterson et al. 2010 |
| MT Com | 0.0830 | 300 | <0.052$v$ | 12.5 | | >11.5 | 12$gs$ | PTK, Szkody et al. 2010 |
| EG Cnc | 0.0600 | 330 | 0.035$s$ | 12.3 | 11.3 | 11.3 | 12.3$gsd$ | Patterson et al. 1998, Szkody et al. 2002, 2010 |
| SDSS 1035+05 | 0.0570 | 170 | 0.055$e$ | 14.0 | | >10.4 | 10.7$se$ | Littlefair et al. 2006 |
| RX 1050–14 | 0.0615 | 80 | <0.055$v$ | 14.0 | | >10.9 | <12$g$ | Mennickent et al. 2001, PTK |
| WZ Sge | 0.0567 | 43 | 0.045$sv$ | 12.8 | 12.8 | 11.6 | 13.5$s$ | Patterson et al. 2002, P98, Godon et al. 2006, T03 |
| V455 And | 0.0563 | 74 | 0.06$s$ | 12.2 | >10.8 | >10.0 | 10.5$gs$ | Araujo-Betancor et al. 2005, Patterson et al. 2010, G09 |
| GW Lib | 0.0533 | 100 | 0.056$s$ | 12.8 | 10.6 | 11.4 | 13.2$s$ | Copperwheat et al. 2009, Vican et al. 2010 |
| PQ And | 0.0558 | 160 | <0.07? | 13.8 | >10.8 | >10.8 | 12.0$s$ | Patterson et al. 2005, Szkody et al. 2010 |
| SDSS 1238–03 | 0.0559 | 140 | | 12.8 | >10.4 | >10.4 | | Zharikov et al. 2006, Aviles et al. 2010 |
| SDSS 1216+05 | 0.0686 | 400 | <0.06$v$ | 12.3 | | >10.0 | <12$g$ | Southworth et al. 2008 |
| BW Scl | 0.0543 | 110 | | 12.3 | | >11.5 | 14.6$s$ | Uthas et al. 2010, Gansicke et al. 2005 |
| V592 Her | 0.0561 | 390 | 0.037?$v$ | 13.7 | | 12.8 | | K09, CBA, Mennickent et al. 2002, Kato 2002 |
| AL Com | 0.0567 | 380 | 0.060s | 13.6 | 10.6 | 10.9 | <15g | Patterson et al. 1996, Szkody et al. 2003 |
| SDSS 0804+51 | 0.0590 | 140 | 0.06 | 11.8 | >10.2 | | <11$g$ | Zharikov et al. 2008, Pavlenko et al. 2006 |
| RX 0232–37 | ~0.066 | 160 | | 12.2 | | >10.2 | <13.5$g$ | K09, CBA |
| SDSS 1501+55 | 0.0568 | 320 | 0.067$e$ | 13.0 | | >10.1 | 12.5$ge$ | Littlefair et al. 2008 |
| ASAS 1536–08 | 0.0641 | 140 | 0.065?$s$ | 11.8 | >11.4 | >11.4 | | PTK |
| OT 1112–35 | 0.0585 | 300 | 0.045?$s$ | 13.4 | | >10.3 | | K09, CBA |
| SDSS 1514+45 | ? | 400 | | 12.8 | | | 10.5$s$ | Dillon et al. 2008, Szkody et al. 2010 |
| OT 0238+35 | 0.0532 | 800 | 0.04?$s$ | >12.0 | | | | K09, CBA, Chochol et al. 2009 |
| UZ Boo | ~0.061 | 270 | | 13.1 | 10.8 | | | CBA, K09 |

NOTES:
1. Type of evidence: $v$ = velocities; $s$ = superhumps; $e$ = eclipses.
2. Accretion component only.
3. Estimate of $<M_{ve}>$ for stars which have had outbursts.
4. Estimate of $<M_{ve}>$ corrected to $i = 57°$. When the only information about outbursts is upper limits, we adopt a "standard outburst": a square wave with $M_v$ = +5.6 and a duration of 18 days.
5. Type of evidence: $s$ = spectroscopy (UV); $e$ = eclipses; $g$ = GALEX color; $d$ = $M_v$ and optical spectra. These are in decreasing order of weight. The entry is "estimated $T_{WD}$ long after outburst". This is complicated by the slowness of the $T_{WD}$ decline after outburst. This has been extensively documented for WZ Sge (Sparks et al. 1993, Slevinsky et al. 1999, Godon et al. 2006) and EG Cnc (Szkody et al. 2002). Since not all outbursts are observed, we tend to adopt the lowest $T_{WD}$ observed, if it is well supported by the data.

TABLE 4 – POPULATION STATISTICS OF DWARF NOVAE

| Distance (pc) | Number | Number with Measured $q$ | Implied* $N_0$ ($\times 10^{-6}$ pc$^{-3}$) | $\langle q \rangle$ |
|---|---|---|---|---|
| <100 | 7 | 7 | 2.0 | 0.079 |
| 100→200 | 25 | 18 | 1.2 | 0.101 |
| 200→300 | 33 | 28 | 0.73 | 0.123 |
| 300→400 | 61 | 32 | 0.77 | 0.136 |
| 400→600 | 65 | 33 | 0.33 | 0.158 |
| 600→1000 | 55 | 33 | 0.17 | 0.161 |
| 1000→1500 | 16 | – | – | |

* Assuming an exponential scale height of 300 pc.

TABLE 5 – THE PERIOD-BOUNCER SCORECARD

| Star | Low $q$ | Cool $T_{wd}$ | Faint $<M_{ve}>$ | Long $P_{orb}$ | Faint $M_{vq}$ | Long $T_{rec}$ | Large DN amplitude | Long DN plateau | Other | Score |
|---|---|---|---|---|---|---|---|---|---|---|
| GD 552 | YY | YYY | Y | YYY | YY | YY | | | Y | 14 |
| EG Cnc | YY | Y | Y | Y | Y | Y | I | Y | | 8 |
| MT Com | Y | Y | | YY | Y | Y | | | | 6 |
| RX 1050–14 | YY | YY | | Y | YY | Y | | | Y | 9 |
| WZ Sge | Y | I | YY | I | Y | Y | Y | Y | | 7 |
| SDSS 1035+05 | YY | YYY | | I | YY | | | | | 7 |
| V455 And | Y | YY | ? | I | Y | I | Y | Y | Y | 7 |
| GW Lib | Y | I | Y | I | Y | Y | Y | Y | | 6 |
| RX 0232–37 | | I | I | YY | Y | | Y | Y | Y | 6 |
| SDSS 1238–03 | | YY | I | I | Y | I | | | Y | 4 |
| PQ And | I | Y | YY | I | Y | Y | Y | Y | | 7 |
| SDSS 1216+05 | YY | Y | YYY | | | | | | | 6 |
| BW Scl | | N? | Y | I | Y | Y | | | Y | 3 |
| V592 Her | Y? | | Y | I | Y | Y | Y | Y | | 5–6 |
| UZ Boo | | | Y | Y | Y | Y | Y | N | | 4? |
| SDSS 0804+51 | Y | YY | I | I | I | ? | | | Y | 4 |
| SDSS 1501+55 | Y | Y | | I | Y | | | | | 3 |
| SDSS 1514+45 | | YYY | | | YY | | | | Y | 6 |
| AL Com | Y | I | Y | I | Y | I | Y | Y | Y | 5 |
| OT 1112–35 | Y? | | I | I | Y | I | Y | Y | | 4 |
| OT 0042+42 | N? | Y | I | I | Y | | Y | Y | | 3 |
| OT 0238+35 | Y? | I | I | I | Y? | I | Y? | Y | | 2–3 |
| V466 And | Y | I | I | I | Y | I | Y | Y | | 4 |
| UW Tri | Y? | I | I | I | Y | I | Y | Y | | 3–4 |
| SDSS 1339+48 | | Y | I | I | Y | I | | | | 2 |
| VX For | | | I | Y | Y | Y | Y | I | | 3–4 |
| OT 0747+06 | | | I | Y | Y | I | Y | Y | | 3–4 |
| SDSS 0902+05 | | | I | N | Y | | Y | Y | | 2–3 |
| ASSA 1536–08 | Y? | | | Y | I | I | N? | N? | Y | 2? |
| ASAS 0025+12 | I | Y | I | I | I | Y | Y | I | | 3 |
| VW Hyi | N | NN | NN | | N | N | N | N | | –9 |
| TY Psc | N | N | N | | N | N | N | N | | –7 |
| YZ Cnc | NN | NN | NN | | NN | N | N | N | | –11 |

A star's score reflects its period-bouncer credentials, although the (descending) order is the author's subjective judgment. Stars which have never erupted are given somewhat lower standing than their score might suggest, since they have never made themselves vulnerable to a test... and since there is always a chance that nonerupters are actually *incorrectly classified* as CVs.

The three stars with negative scores are added for comparison; the majority of catalogued and well-studied dwarf novae have large negative scores.

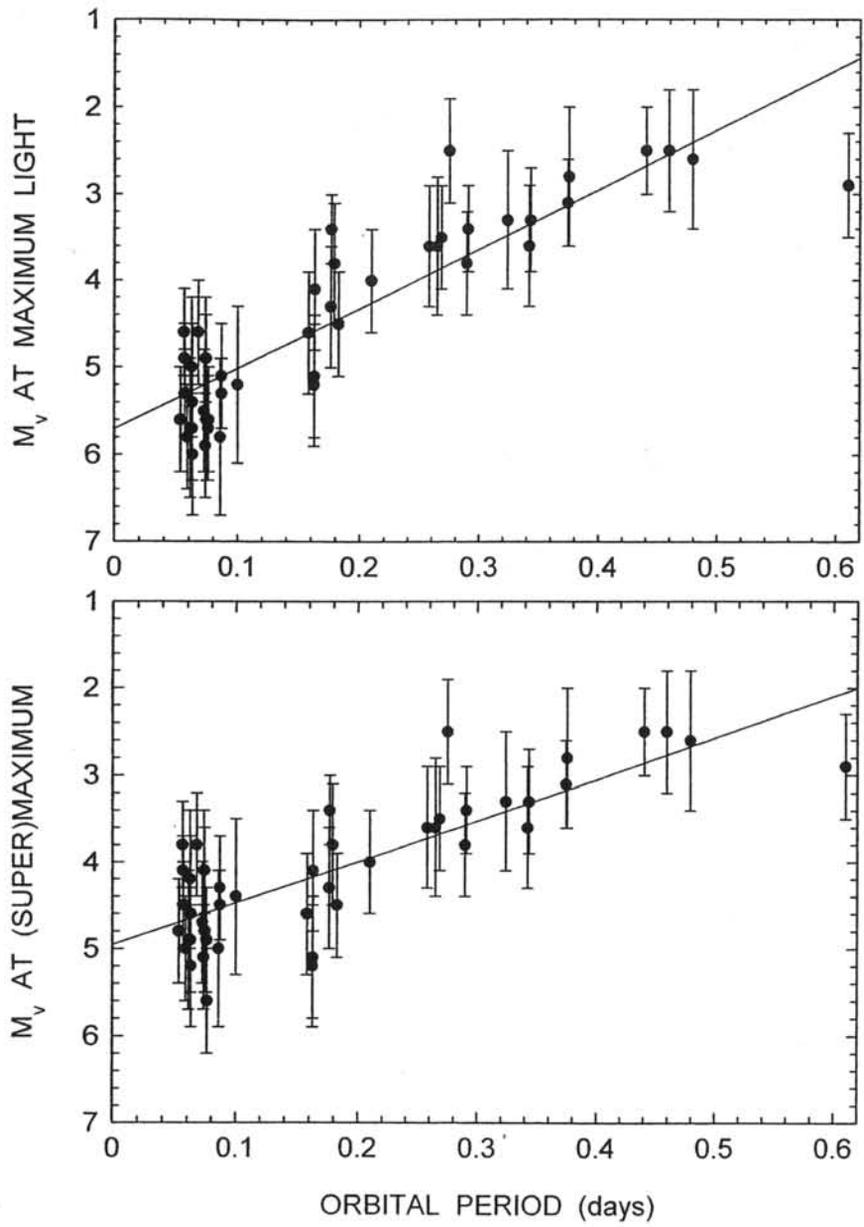

Figure 1.

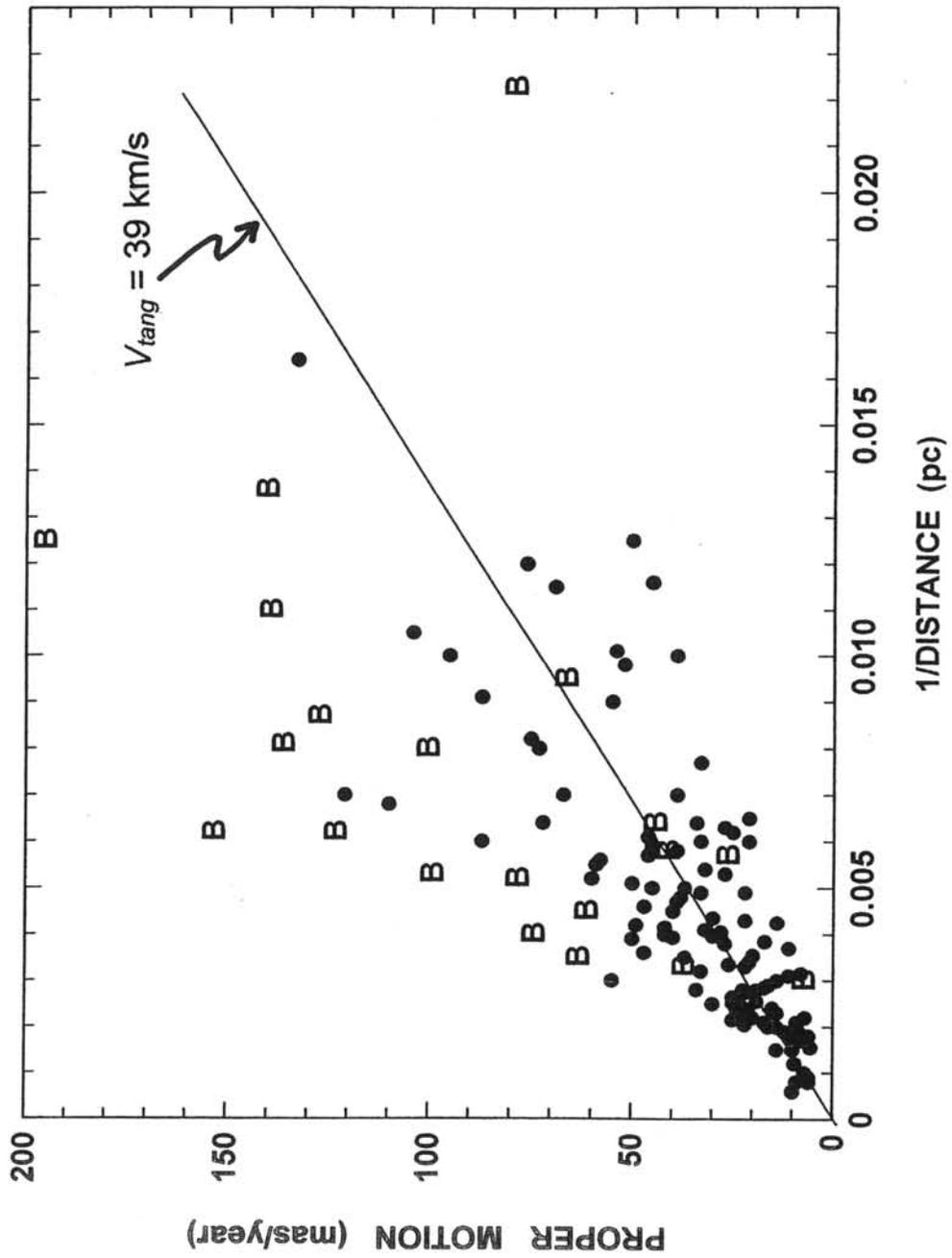

Figure 2.

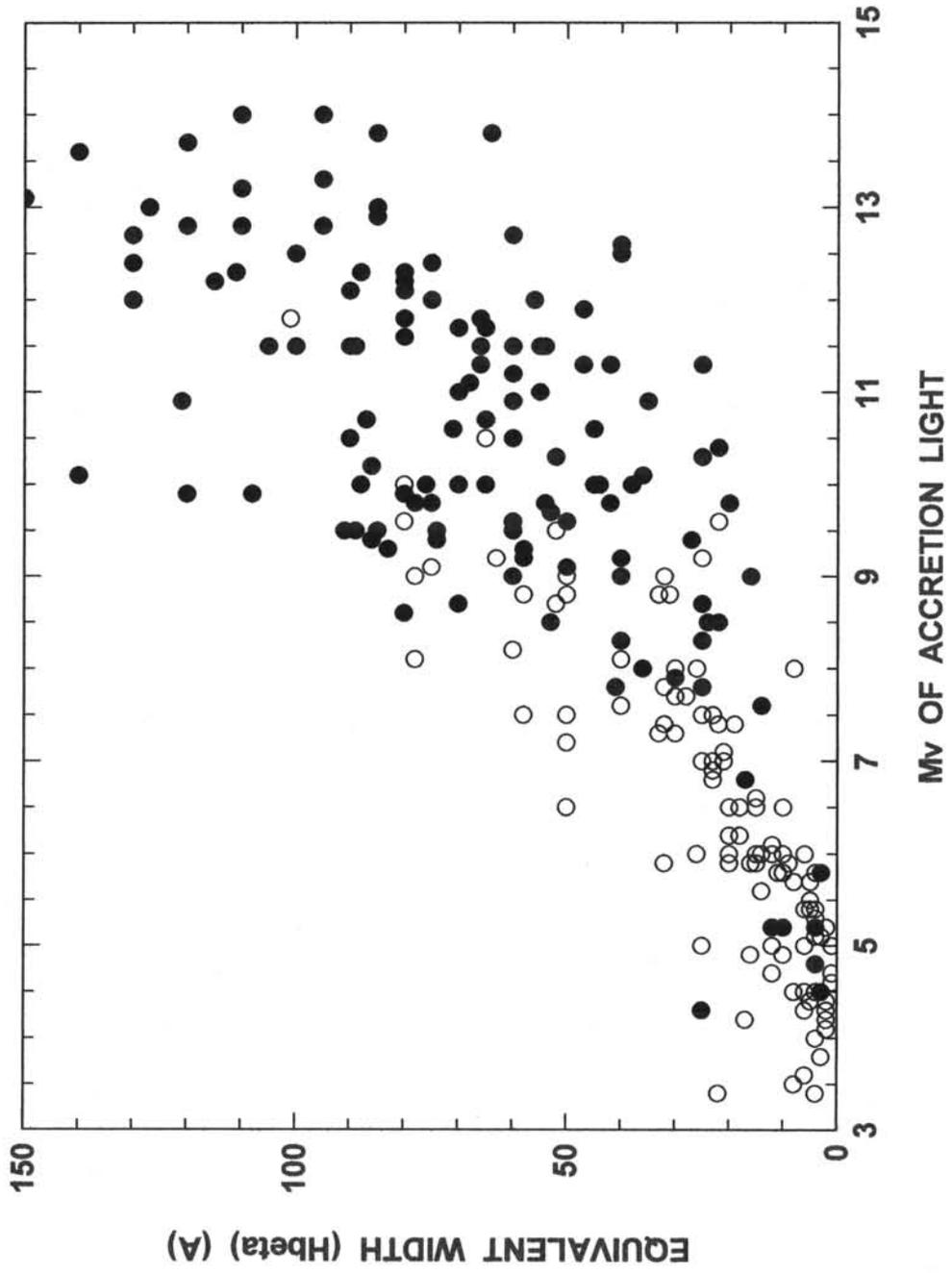

Figure 3.

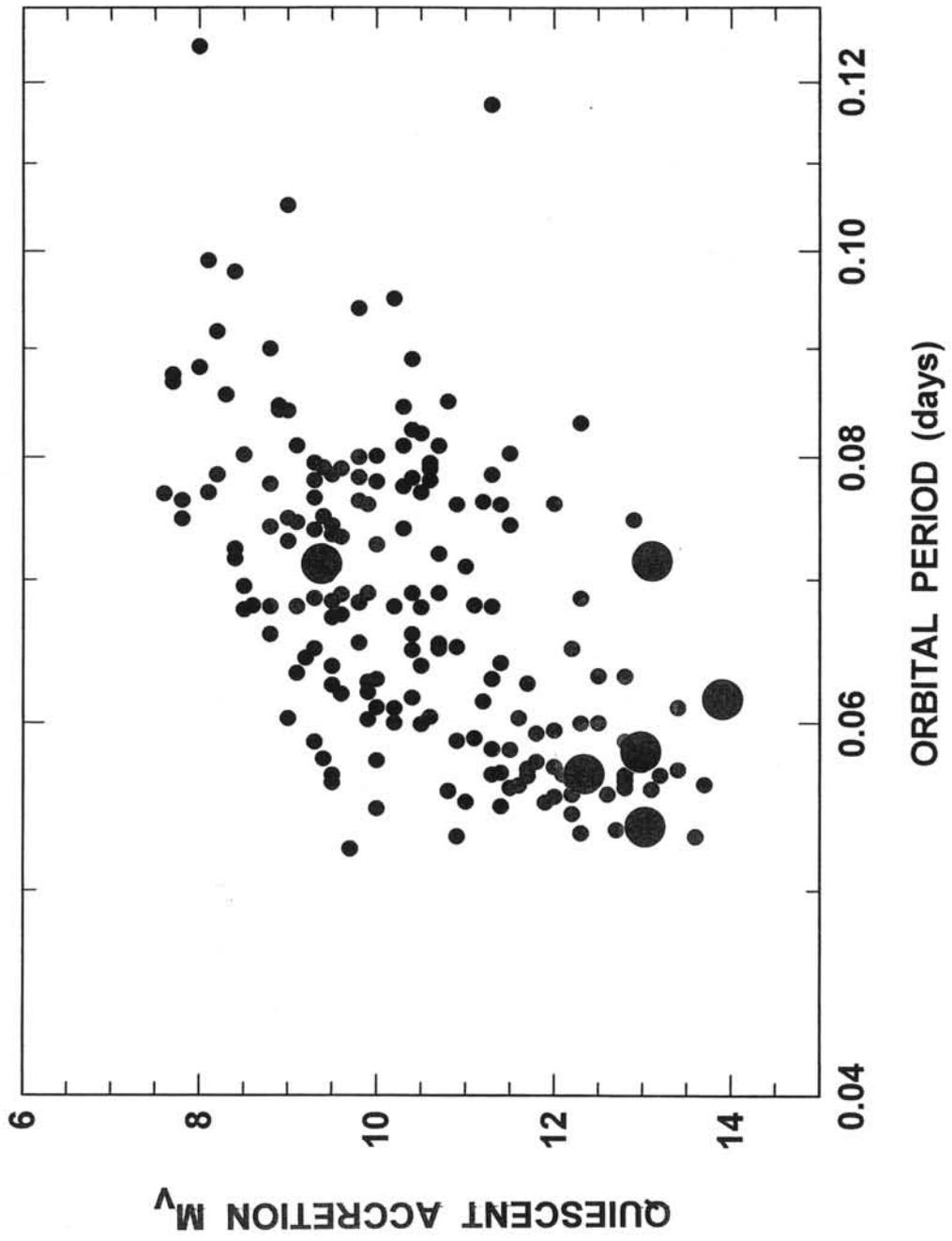

Figure 4.

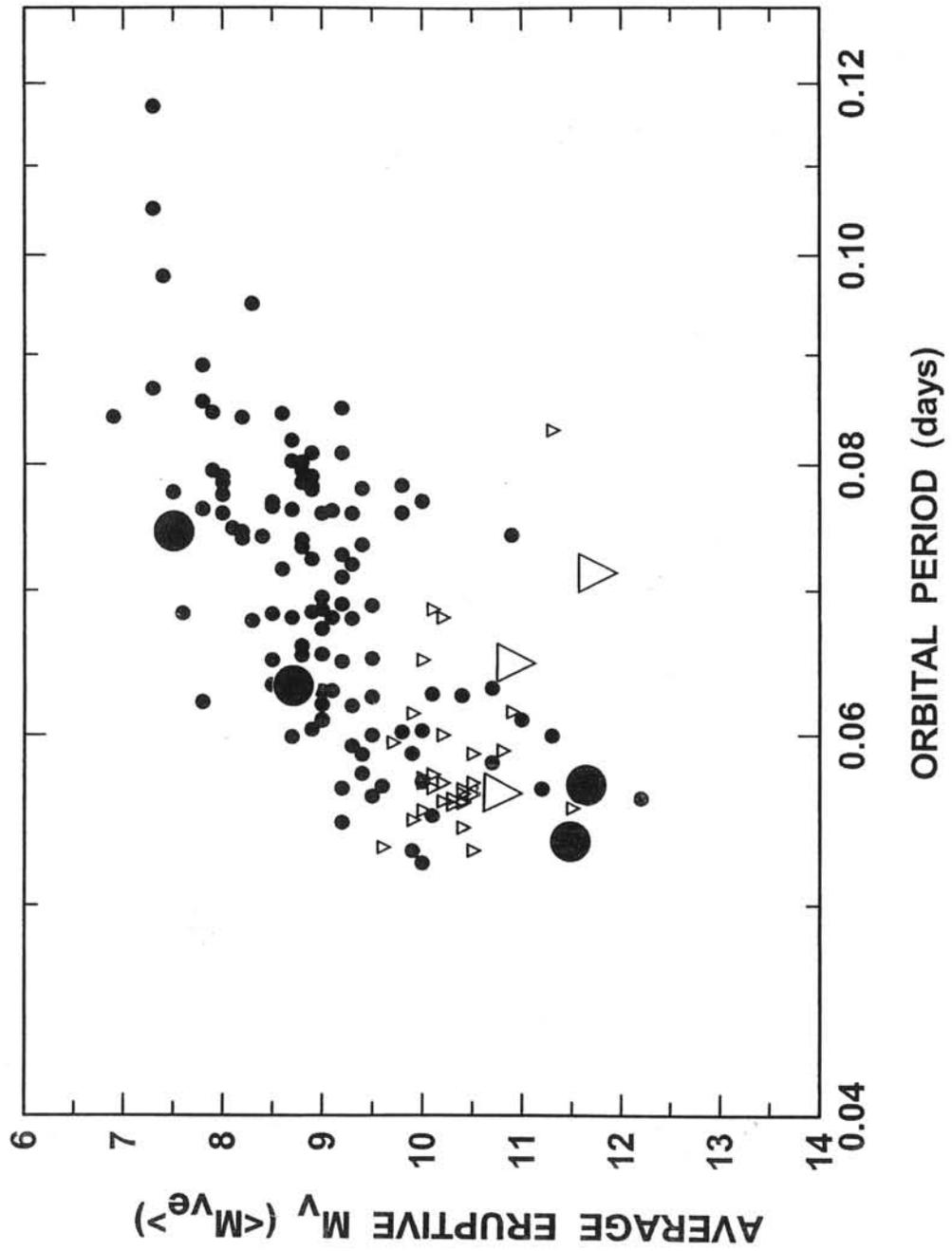

Figure 5.

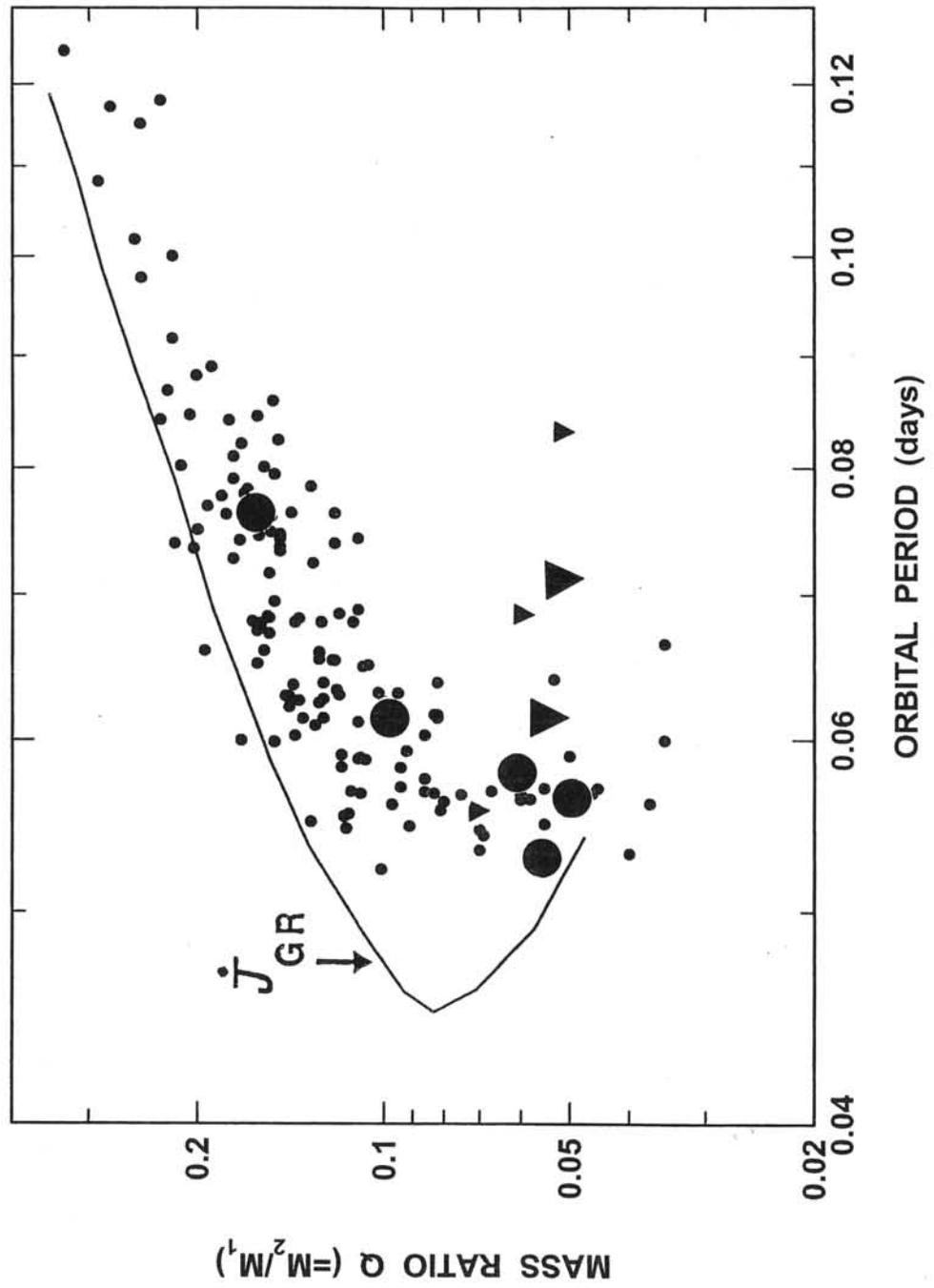

Figure 6.

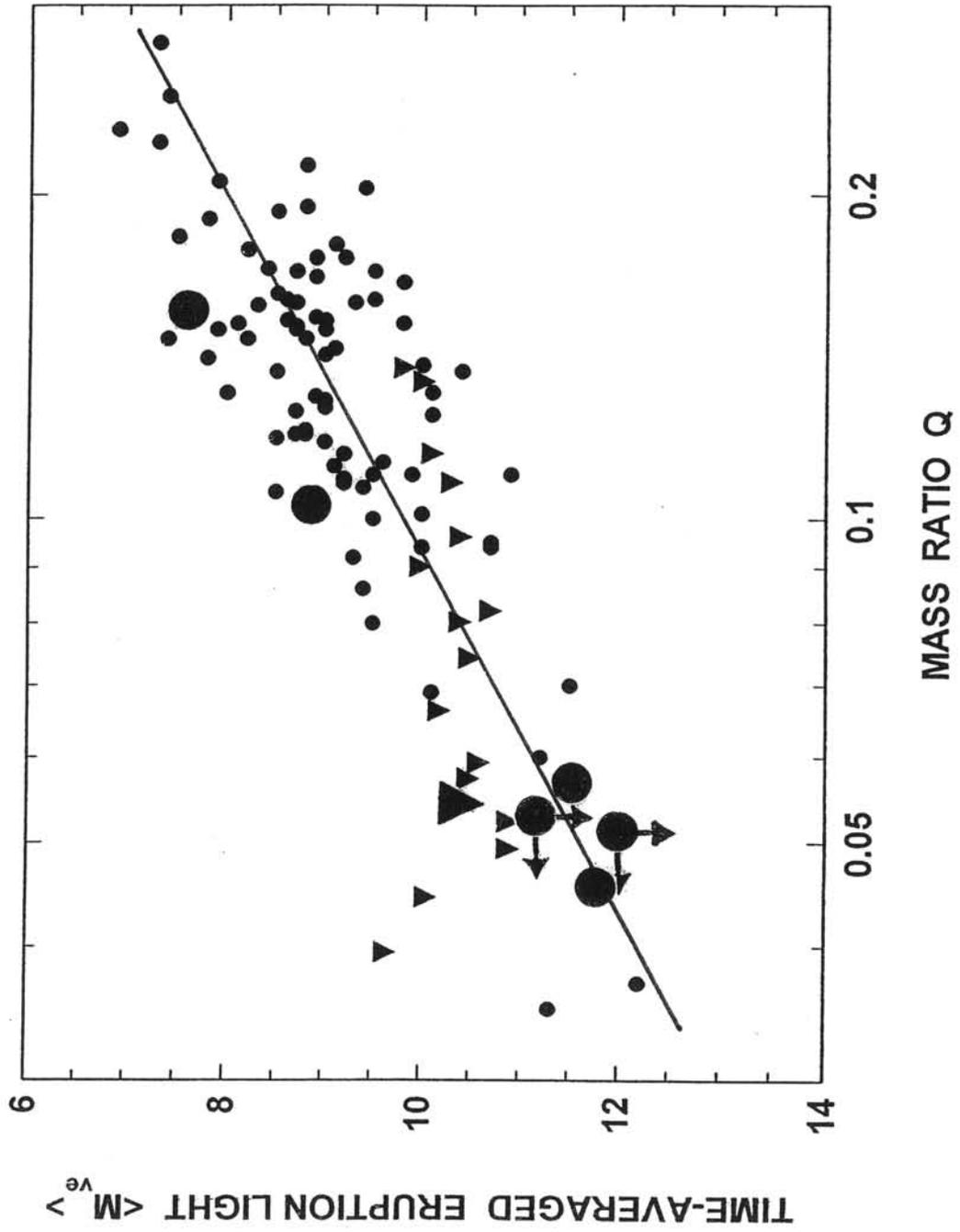

Figure 7.

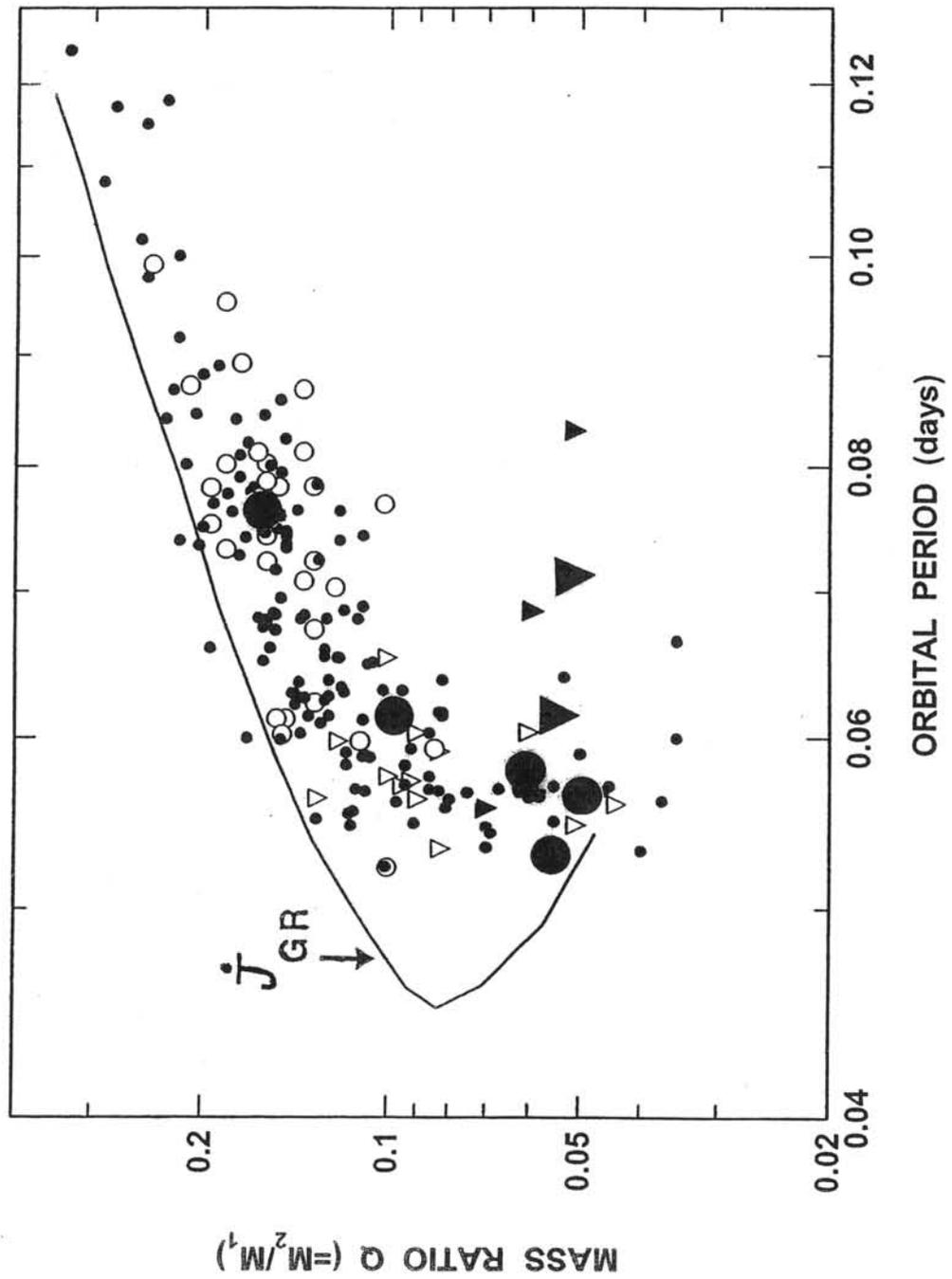

Figure 8.

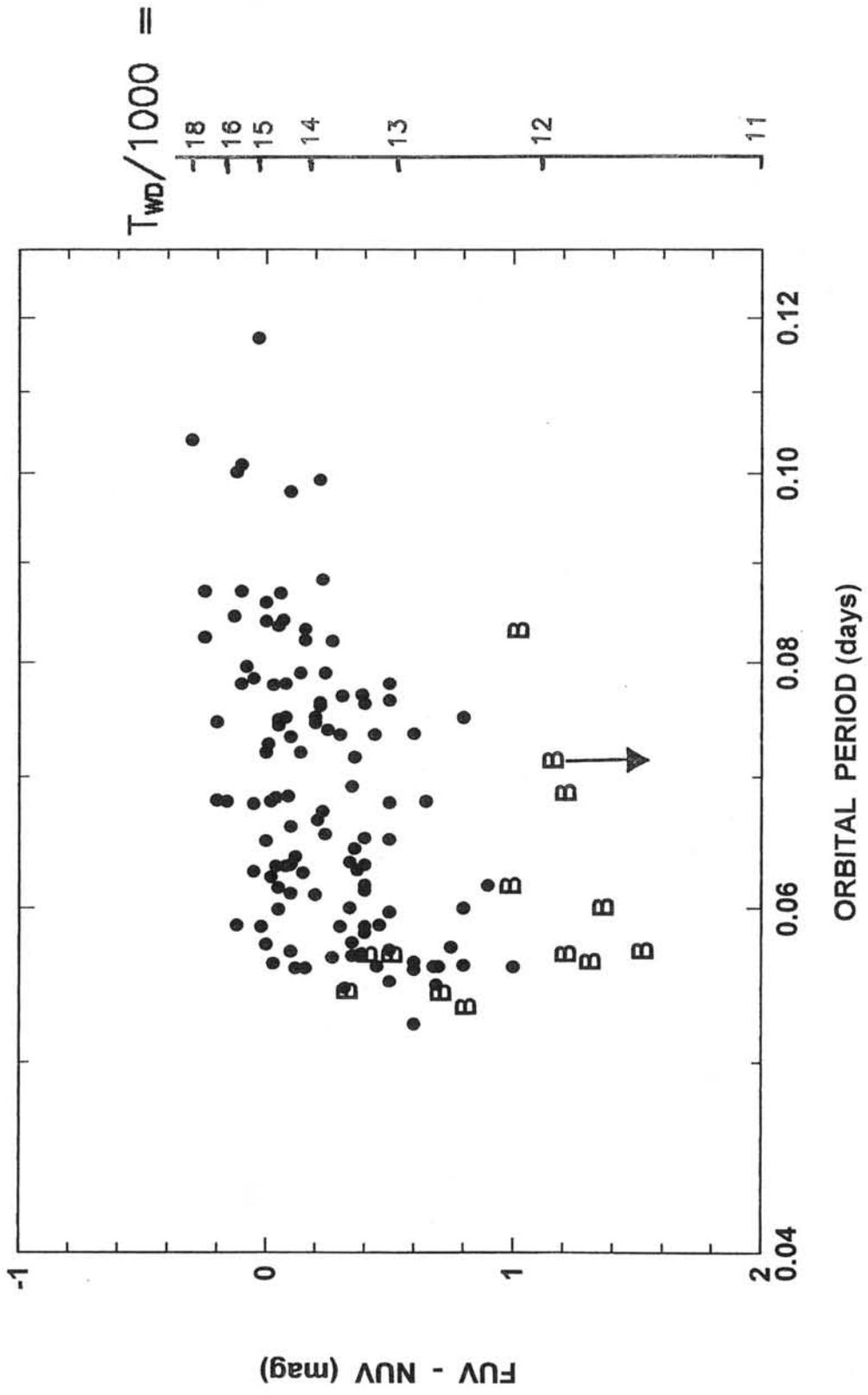

Figure 9.

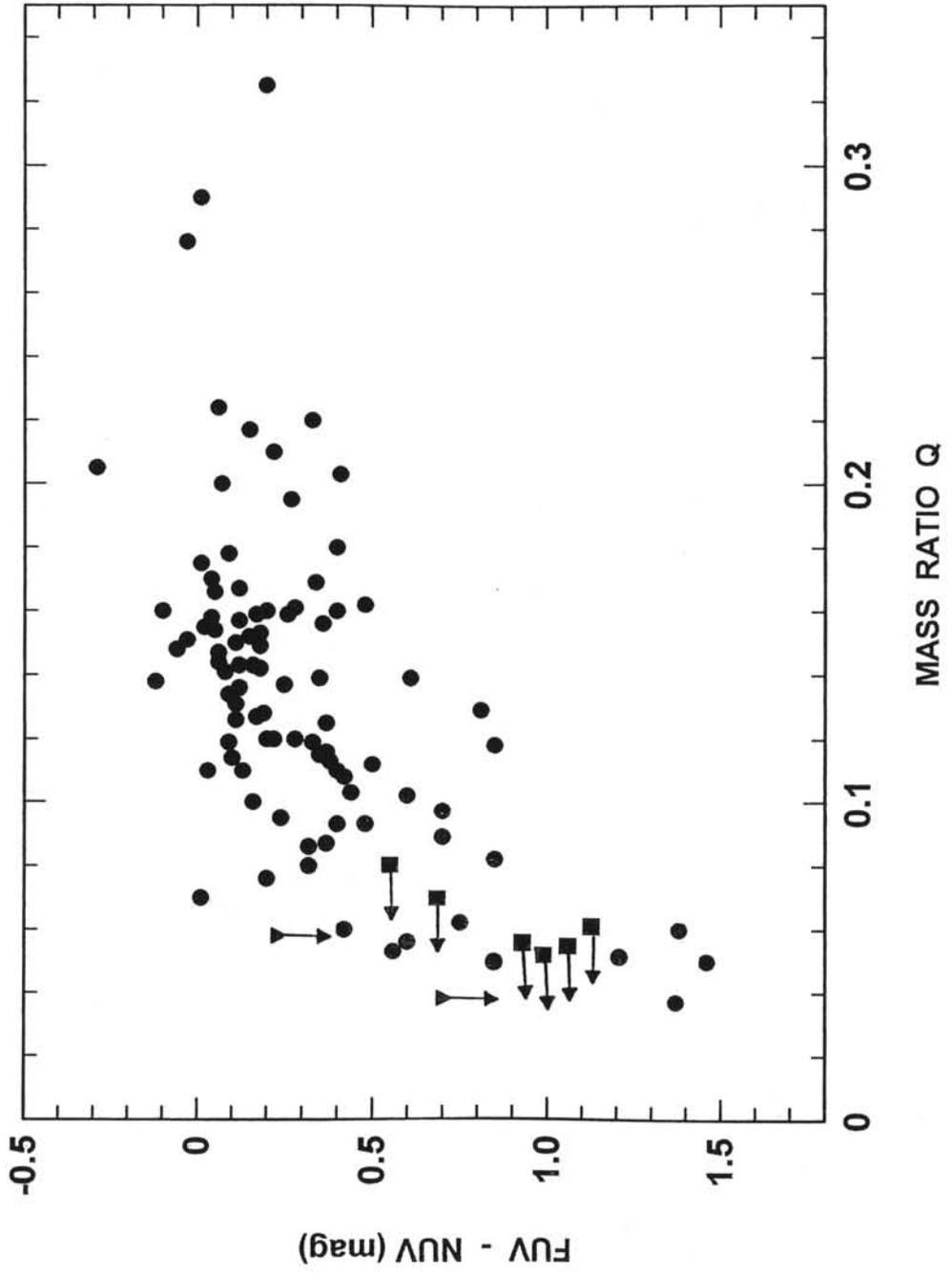

Figure 10.

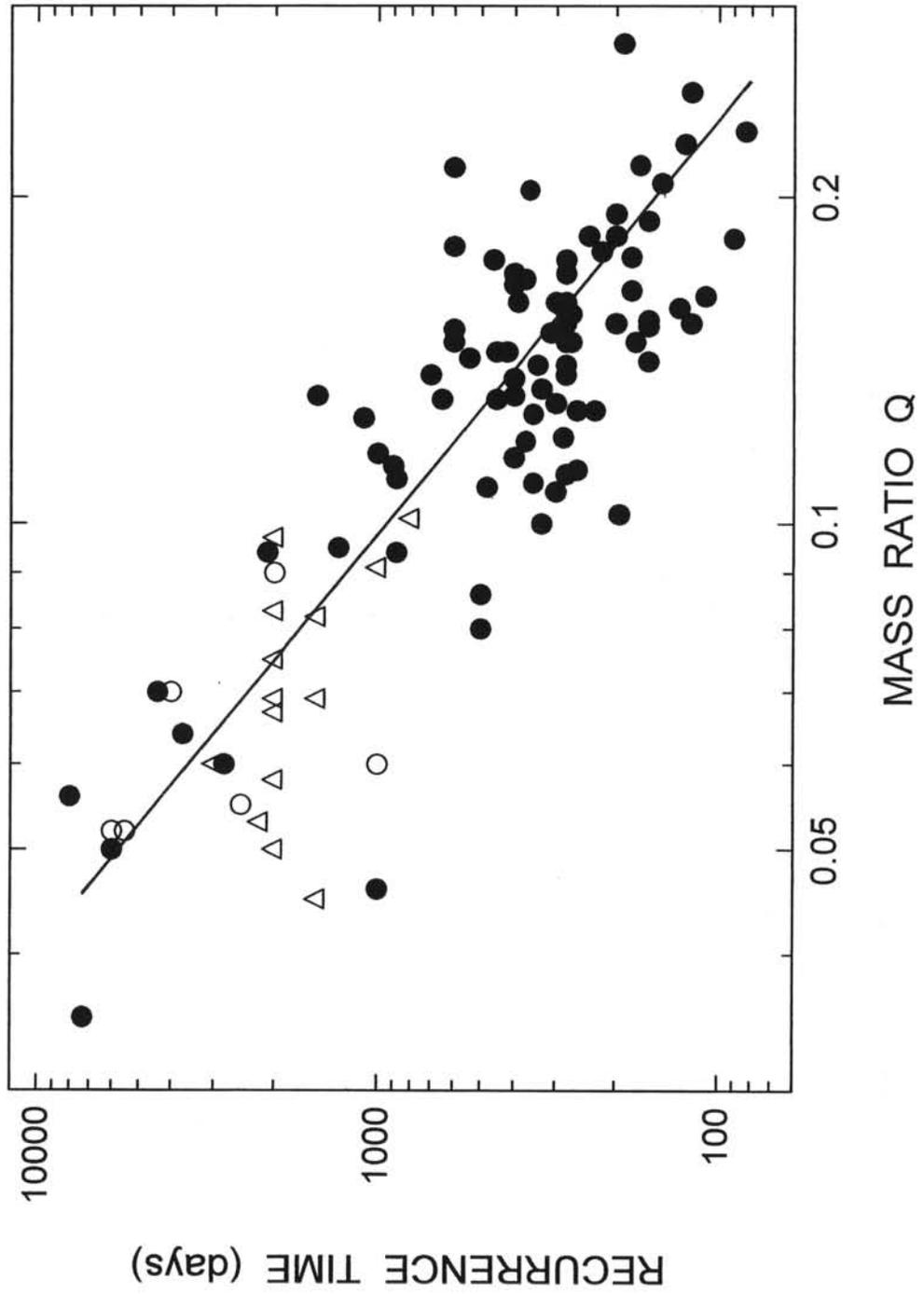

Figure 11.

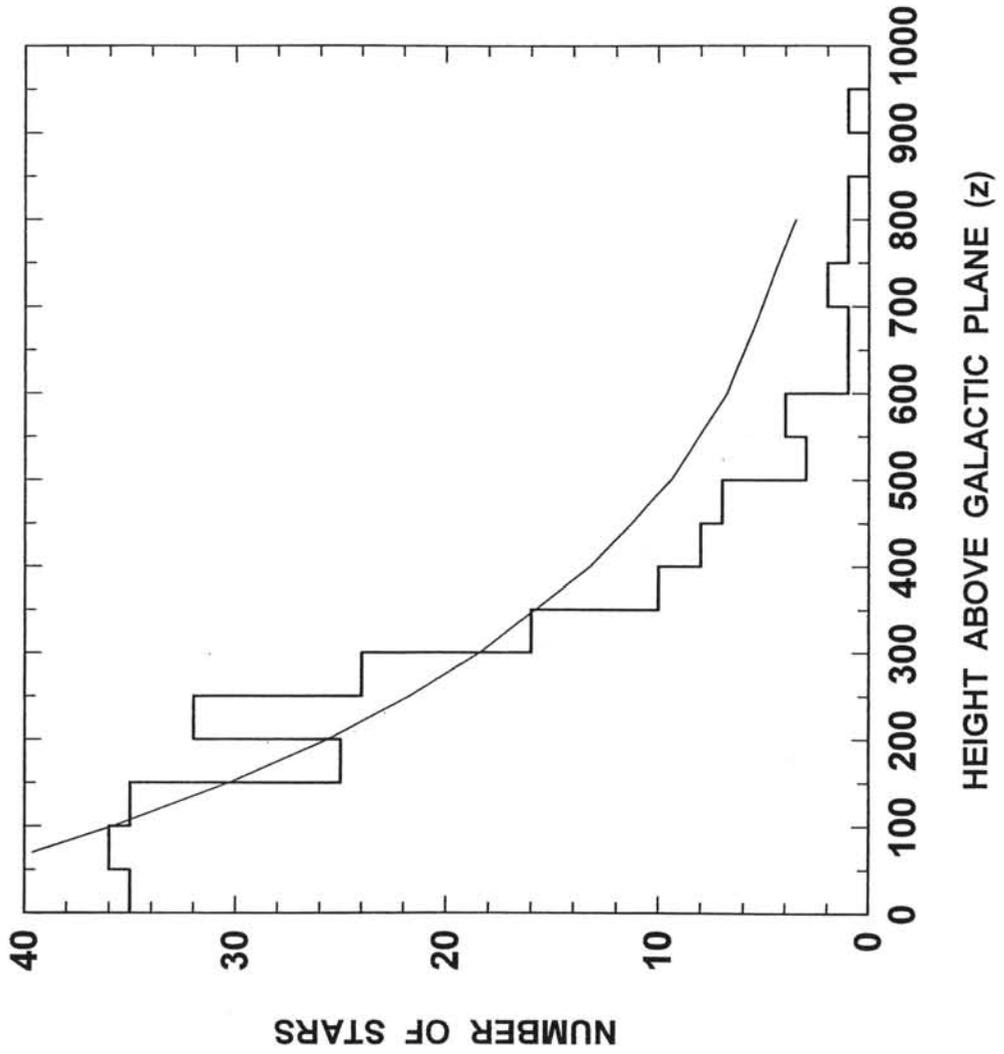

Figure 12.

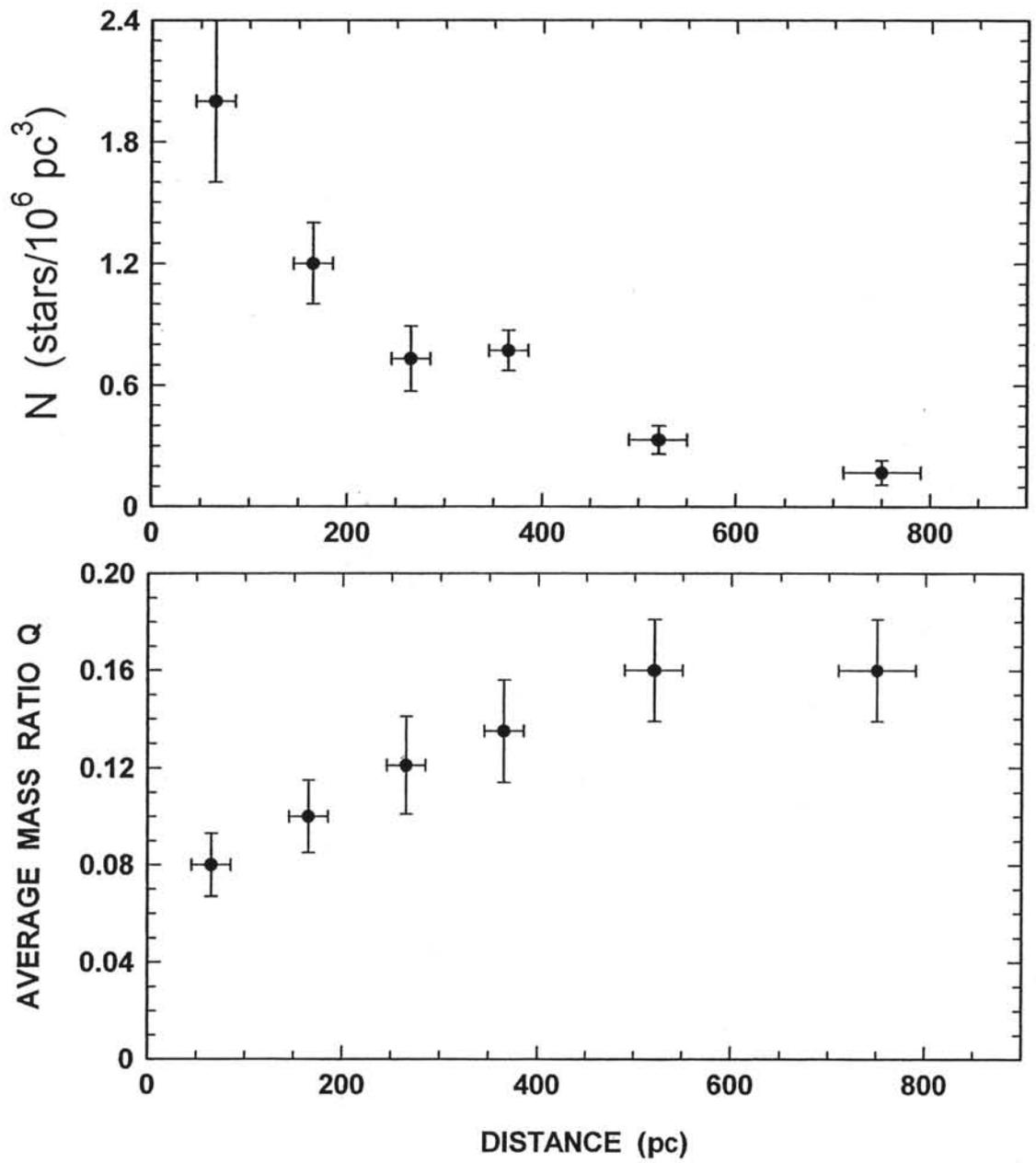

Figure 13.